\documentclass[amsmath,amssymb,aps,prb,twocolumn,longbibliography,superscriptaddress]{revtex4-2}

\usepackage{hyperref}
\usepackage[pdftex,dvipsnames]{xcolor}
\usepackage{bm}
\usepackage{graphicx}
\usepackage{euscript}
\usepackage{appendix}
\usepackage{lipsum}                     
\usepackage{xargs} 
\usepackage[dvipsnames]{xcolor}
\usepackage[normalem]{ulem} % for \sout (strikethrough)
\usepackage{float}
\usepackage{soul}
\usepackage{tikz}
\usetikzlibrary{positioning,calc}

\begin{document}

\title{Hierarchical Reduction of Invariant Subspaces: From Fragmentation to Many-Body Scars}% Force line breaks with \\

\author{Chihiro Matsui}
 \affiliation{Department of Mathematical Sciences, The University of Tokyo, 3-8-1 Komaba, Meguro-ku, Tokyo 153-8914, Japan}%Lines break automatically or can be forced with \\

\date{\today}% It is always \today, today,

\begin{abstract}

We investigate a hierarchy of invariant Hilbert-space structures connecting Hilbert-space fragmentation, partial solvability, and quantum many-body scars. For spin chains with fragmented spin-$1/2$ subspaces and their local deformations, we show, using the spin-$1$ $XXC$ chain as a representative example, how successive local deformations reduce extensive irreducible-string fragmentation through an implicitly embedded spin-$1/2$ invariant space and ultimately to a single scar tower, connecting the $XXC$, perturbed $XXC$, and spin-$1$ $XY$ scar models. 
Through the intermediate embedded space, local dynamics couples states from distinct frozen irreducible-string sectors and selects a new decomposition into representations, from which smaller invariant spaces can be retained. We also establish a rigidity constraint on isolating nonsymmetric towers and a constructive correspondence that transfers a reduction between locally realizable representations of the embedded dynamics. The latter demonstrates that the reduction hierarchy can be transferred through a unitary transformation within the embedded subspace, yielding an exact scar tower built on a N\'{e}el reference state without requiring unitary equivalence of the full Hamiltonians. 

\end{abstract}

\maketitle

%%%%%%%%%%%%%%%%%%%%%%%%%%%%%%%%%%%%%%%%%%%%%%%%%%%%%%%
%%%%%%%%%%%%%%%%%%%%%%%%%%%%%%%%%%%%%%%%%%%%%%%%%%%%%%%
\section{Introduction}
\label{sec:intro}
%%%%%%%%%%%%%%%%%%%%%%%%%%%%%%%%%%%%%%%%%%%%%%%%%%%%%%%
%%%%%%%%%%%%%%%%%%%%%%%%%%%%%%%%%%%%%%%%%%%%%%%%%%%%%%%

The boundary between thermalization and exact solvability is considerably richer than the conventional dichotomy between nonintegrable and integrable quantum systems. Generic interacting systems are expected to thermalize according to the eigenstate thermalization hypothesis (ETH)~\cite{bib:D1991,bib:S1994,bib:S1999,bib:RDO2008,bib:DKPR2016}, whereas integrable systems evade this mechanism through an extensive set of conserved quantities. A third possibility is that exact or otherwise constrained dynamics survives only within a proper invariant subspace of an otherwise nonintegrable Hamiltonian. Such partially solvable systems are useful not only as exceptions to ETH, but also as controlled settings for asking a more structural question: how far can an invariant Hilbert-space structure be reduced under local deformations, and what larger structures must necessarily survive in order to protect a prescribed nonthermal subspace?

Hilbert-space fragmentation (HSF)~\cite{bib:P2022,bib:MBR2022,bib:KHN2020,bib:SRVKP2020}, isolated solvable sectors, and quantum many-body scars (QMBS)~\cite{bib:MRBR2018,bib:MRB2018,bib:SI2019,bib:ISX2019,bib:OCMCCN2019,bib:CIKM2023} provide distinct reference points for this question. In a fragmented system, the local Hamiltonian decomposes the Hilbert space into many dynamically disconnected sectors, often exponentially many in the system size, that cannot in general be distinguished by conventional global symmetries. In a partially solvable system~\cite{bib:S1995,bib:SA1996,bib:M2024,bib:ADS2024,bib:KMPP2025}, only selected extended sectors retain closed dynamics, while the complementary Hilbert space may be nonintegrable or thermalizing. QMBS represent a still more reduced structure, consisting of a sparse set of atypical eigenstates embedded in an otherwise thermal spectrum and potentially generating coherent revivals from simple initial states~\cite{bib:TMASP2018,bib:SAP2021}. Exact scar towers are frequently stabilized by local annihilation conditions or restricted spectrum-generating algebras~\cite{bib:MRB2020}. These phenomena can therefore be viewed as different levels of a hierarchy distinguished by how much invariant Hilbert-space structure survives. The intermediate level is not merely a larger container for scar states: dynamics introduced there can connect states carrying different fragmentation labels and organize them into a new representation-based decomposition.

Connections between HSF and QMBS have been discussed in several contexts~\cite{bib:LX2021,bib:YMP2026}. In particular, a recent work~\cite{bib:YMP2026} demonstrated a direct connection by partially activating kinetically constrained fragmented structures to generate scarred dynamics.
Here we take a different perspective. Rather than regarding fragmentation and scars themselves as the primary objects, we ask how the underlying invariant subspaces are successively reduced under local deformations: which structures can be removed while a prescribed smaller subspace remains invariant, and when does the attempt to preserve that subspace necessarily restore a larger one? This question is particularly sharp for scar towers built on nonsymmetric reference states. Their preservation is a priori a much weaker requirement than preservation of an extended invariant sector, yet locality and symmetry may prevent the two from being separated. 

In this work, we develop a representation-based construction of invariant-subspace hierarchies for spin chains exhibiting irreducible-string fragmentation and their local deformations. The spin-$1$ $XXC$ chain~\cite{bib:M1998,bib:M1999} provides a representative example. The $XXC$ Hamiltonian acts as a scalar on its exponentially large embedded space $W_{\rm emb}$, which contains states from distinct frozen irreducible-string (IS) sectors. Introducing local dynamics in this space can connect these sectors, replacing part of the original IS-based fragmentation by a decomposition determined by the embedded spin-$1/2$ dynamics. This first reduction leads to the perturbed $XXC$ model~\cite{bib:M2024,bib:KMPP2025}, in which $W_{\rm emb}$ remains invariant. A subsequent local coupling to the complement can preserve a selected polynomial-dimensional representation subspace while breaking the invariance of $W_{\rm emb}$ itself, as realized in the spin-$1$ $XY$ scar model~\cite{bib:SI2019}. These models thus provide successive levels of a hierarchy of invariant-subspace reductions. 

We establish two complementary results within this framework. First, for translationally invariant, $U(1)$-conserving nearest-neighbor perturbations, preserving a nonsymmetric $\mathfrak{su}(2)$ tower of nonmaximal spin forces the larger embedded space to remain invariant. Nonmaximal-spin scar towers are nevertheless possible beyond these constraints, as demonstrated in \cite{bib:OBCK2020}. Second, a unitary transformation within $W_{\rm emb}$ can relate different realizations of the embedded dynamics. When the corresponding leakage operators can also be transformed locally, the reduction to a smaller invariant subspace can be transferred between these realizations. We demonstrate this correspondence through the staggered transformation between polarized and alternating configurations, which yields an exact N\'{e}el-reference scar tower. The resulting full spin-$1$ Hamiltonians need not be unitarily equivalent.

\begin{figure*}[t]
\centering
\begin{tikzpicture}[
    >=stealth,
    box/.style={
        draw,
        rounded corners=3pt,
        align=center,
        minimum height=1.05cm,
        text width=6.0cm,
        inner sep=5pt
    },
    arrow/.style={
        ->,
        line width=0.7pt
    },
    node distance=1.15cm and 1.6cm
]

%------------------------------------------------
% Column labels
%------------------------------------------------
\node (poltitle) {\textbf{Polarized realization}};
\node[right=5.7cm of poltitle] (neeltitle)
    {\textbf{Alternating / N\'eel realization}};

%------------------------------------------------
% Fragmentation level
%------------------------------------------------
\node[box, below=0.18cm of poltitle] (fragpol) {
    \textbf{Fragmentation}\\
    XXC model: polarized IS
};

\node[box, below=0.18cm of neeltitle] (fragneel) {
    \textbf{Fragmentation}\\
    XXC model: alternating IS
};

%------------------------------------------------
% Embedded level
%------------------------------------------------
\node[box, below=of fragpol] (embpol) {
    \textbf{Embedded spin-$1/2$ invariant space}\\
    perturbed XXC model: $W_{\rm emb}$
};

\node[box, below=of fragneel] (embneel) {
    \textbf{Embedded invariant structure}\\
    perturbed XXC model: $W_{\rm emb}$
};

%------------------------------------------------
% Scar level
%------------------------------------------------
\node[box, below=of embpol] (scarpol) {
    \textbf{Single invariant tower}\\
    polarized scar tower: $W_{\rm scar}$
};

\node[box, below=of embneel] (scarneel) {
    \textbf{Single invariant tower}\\
    N\'eel scar tower: $W'_{\rm scar}$
};

%------------------------------------------------
% Vertical reductions
%------------------------------------------------
\draw[arrow] (fragpol) --
    node[right, align=left, font=\small]
    {break IS fragmentation\\ preserve $W_{\rm emb}$}
    (embpol);

\draw[arrow] (embpol) --
    node[right, align=left, font=\small]
    {break $W_{\rm emb}$\\ preserve one tower}
    (scarpol);

\draw[arrow] (fragneel) -- (embneel);
\draw[arrow] (embneel) -- (scarneel);

%------------------------------------------------
% Horizontal unitary transformations
%------------------------------------------------
\draw[arrow] (fragpol) --
    node[above, font=\small] {IS-sector mapping}
    (fragneel);

\draw[arrow] (embpol) --
    node[above, font=\small] {staggered unitary}
    (embneel);

\draw[arrow] (scarpol) --
    node[above, font=\small] {staggered unitary}
    (scarneel);

\end{tikzpicture}

\caption{
Schematic hierarchy of invariant-subspace reduction and its
staggered-unitary realization.
Successive local deformations reduce the invariant structure from
IS fragmentation to the embedded spin-$1/2$ space $W_{\rm emb}$ and
finally to a single scar tower.
Local embedded dynamics couples states with different frozen IS labels and reorganizes them into representation subspaces. The staggered transformation relates the restricted embedded dynamics and the selected towers in the two realizations; full-Hilbert-space unitary equivalence is not assumed.}
\label{fig:hierarchy}
\end{figure*}
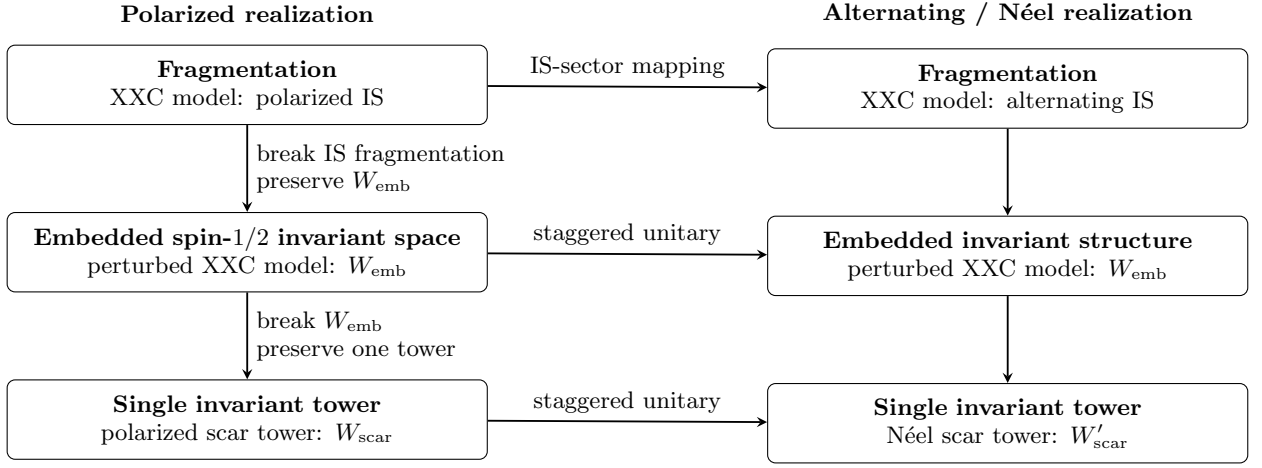

The remainder of this paper is organized as follows. 
In Sec.~II, we review the IS fragmentation of the $XXC$ model and introduce its implicitly embedded spin-$1/2$ invariant structure. 
In Sec.~III, we construct the partially solvable level of the hierarchy by introducing perturbations within the implicitly embedded spin-$1/2$ space. 
In Sec.~IV, we reduce this structure further to the scar level and establish the rigidity of nonsymmetric $\mathfrak{su}(2)$ towers. 
In Sec.~V, we explain how a unitary transformation within $W_{\rm emb}$ can be extended to the subsequent reduction through a local transformation of the leakage operators, and illustrate this correspondence with polarized and alternating IS configurations. 
In Sec.~VI, we carry these transformations through the hierarchy and obtain the N\'{e}el scar construction. 
We conclude in Sec.~VII. Technical details are provided in the appendices.

%%%%%%%%%%%%%%%%%%%%%%%%%%%%%%%%%%%%%%%%%%%%%%%%%%%%%%%
%%%%%%%%%%%%%%%%%%%%%%%%%%%%%%%%%%%%%%%%%%%%%%%%%%%%%%%
\section{Implicit spin-$1/2$ invariant subspace of the $XXC$ model}
\label{sec:XXC}
%%%%%%%%%%%%%%%%%%%%%%%%%%%%%%%%%%%%%%%%%%%%%%%%%%%%%%%
%%%%%%%%%%%%%%%%%%%%%%%%%%%%%%%%%%%%%%%%%%%%%%%%%%%%%%%

We begin with the open $XXC$ model~\cite{bib:M1998,bib:AM1998}, which provides the fragmented limit of the hierarchy discussed in this work. Besides its well-known HSF~\cite{bib:KMPP2025}, we point out that the model contains an implicit spin-$1/2$ invariant subspace that serves as the starting point for all subsequent constructions. 

Consider a spin-$\ell/2$ chain on $(\mathbb C^{\ell+1})^{\otimes L} = {\rm span}\{ |0\rangle,\ldots,|\ell\rangle \}^{\otimes L}$, whose nearest-neighbor Hamiltonian is
\begin{align}
    H^{\rm gen}_{XXC}(a,b,c)
    = \sum_{j=1}^{L-1} (h^{\rm gen}_{XXC}(a,b,c))_{j,j+1},
\end{align}
with local interaction
\begin{align}
    h^{\rm gen}_{XXC}(a,b,c)
    =& a\Big( \sum_{s,s'\in A} |ss'\rangle\langle ss'|
    +
    \sum_{t,t'\in B} |tt'\rangle\langle tt'| \Big) \nonumber \\
    &+ b\sum_{s\in A,\, t\in B}\Big( |ts\rangle\langle ts| + |st\rangle\langle st| \Big) \nonumber \\
    &+ c\sum_{s\in A,\, t\in B} \Big( |ts\rangle\langle st| + |st\rangle\langle ts| \Big), \label{eq:gen_XXC}
\end{align}
where $A\cup B = \{0,1,\ldots,\ell\}$ and $A\cap B = \varnothing$. Since the local interactions \eqref{eq:gen_XXC} never exchange species in $A$ and $B$, the Hilbert space is fragmented according to the initial order of species within $A$ and $B$, respectively. 

Throughout the main text, we consider the minimal spin-$1$ realization of this broader class of $XXC$ model by specializing 
\begin{align}
    A=\{0,2\}, \qquad B=\{1\},
\end{align}
which reduces \eqref{eq:gen_XXC} to the spin-$1$ Hamiltonian given by 
\begin{align}
    h_{XXC}(a,b,c)
    &= a \big( \sum_{s,t=0,2} |st \rangle \langle st| + |11 \rangle \langle 11| \big) \nonumber \\
    &+ b \sum_{s=0,2} \big( |s1 \rangle \langle s1| + |1s \rangle \langle 1s| \big) \nonumber \\
    &+ c \sum_{s=0,2} \big( |s1 \rangle \langle 1s| + |1s \rangle \langle s1| \big) \\
    &= c (S^x \otimes S^x + S^y \otimes S^y) \nonumber \\
    &+ (b-a) \big( (S^z)^2 \otimes \bm{1} + \bm{1} \otimes (S^z)^2  \nonumber \\
    &\hspace{18mm}- 2(S^z)^2 \otimes (S^z)^2 \big) \nonumber \\
    &+ a\bm 1. \label{eq:XXC}
\end{align}
The general $XXC$ family is revisited in Appendix~\ref{app:general_xxc}, where we discuss which steps of the reduction hierarchy can extend beyond the spin-$1$ realization. 

This model is known to evade thermalization for any choice of initial state due to fragmentation of the Hilbert space into exponentially many invariant subspaces~\cite{bib:SRVKP2020,bib:MBR2022,bib:KMPP2025}. 
The fragmentation is brought about by frozen partial configurations~\cite{bib:KMPP2025}, each represented by a word $\lambda\in\{0,2\}^{\ell_\lambda}$, where $\ell_{\lambda}$ denotes its length. The frozen word $\lambda$ is called an ``irreducible string'' (IS)~\cite{bib:DB1993,bib:BD1994,bib:MBD1997}, which uniquely labels each fragmented subspace $W_{\lambda}$.

Besides exhibiting fragmentation, the dynamics within each fragmented IS-labeled subspace is integrable and effectively described by the spin-$1/2$ $XXZ$ model with fixed magnetization $M_z = \ell_{\lambda}-L/2$. Using the projector $P_{M_z}$ onto the sector with total magnetization $M_z$, we have
\begin{align}
    H_{XXC} \big|_{W_\lambda} 
    &= P_{\ell_{\lambda}-L/2} H_{XXZ} P_{\ell_{\lambda}-L/2} \nonumber \\
    &= \sum_{j=1}^{L-1} P_{\ell_{\lambda}-L/2} (h_{XXZ})_{j,j+1} P_{\ell_{\lambda}-L/2}, 
\end{align}
where 
\begin{align}
    h_{XXZ}(a,b,c) 
    &= a \big( |\uparrow\uparrow \rangle \langle \uparrow\uparrow| + |\downarrow\downarrow \rangle \langle \downarrow\downarrow| \big) \nonumber \\
    &+ b \big( |\uparrow\downarrow \rangle \langle \uparrow\downarrow| + |\downarrow\uparrow \rangle \langle \downarrow\uparrow| \big) \nonumber \\
    &+ c \big( |\uparrow\downarrow \rangle \langle \downarrow\uparrow| + |\downarrow\uparrow \rangle \langle \uparrow\downarrow| \big) \label{eq:h_XXZ} \\
    &= \frac{c}{2} ( \sigma^x \otimes \sigma^x + \sigma^y \otimes \sigma^y ) \nonumber \\
    &+ \frac{a-b}{2} \sigma^z \otimes \sigma^z 
    + \frac{a+b}{2} \bm{1},
\end{align}
independently of the word $\lambda$. This IS-independent effective dynamics is the IS-shadow description: although it retains the mobile-particle pattern, it does not distinguish the internal $0/2$ sequence carried by an IS. The microscopic IS word remains conserved under the unperturbed $XXC$ Hamiltonian. 
This integrability can also be established by the existence of an $R$-matrix solving the Yang--Baxter equation~\cite{bib:M1998,bib:AM1998}. 

Besides the IS-based fragmentation discussed above, the $XXC$ model possesses another invariant structure that will play a central role below. Independently of the IS decomposition, the subspace $W_{\rm emb} = {\rm span}\{|0\rangle,|2\rangle\}^{\otimes L}$ is invariant under the $XXC$ Hamiltonian. Within this subspace, the Hamiltonian acts simply as a scalar,
\begin{align} \label{eq:scalar}
    H_{XXC}(a,b,c) \big|_{W_{\rm emb}} = a(L-1) \bm{1}. 
\end{align}
Thus, $W_{\rm emb}$ constitutes an implicit spin-$1/2$ invariant subspace embedded in the spin-$1$ Hilbert space. The scalar action leaves the relative dynamics of its different frozen IS words undetermined by $H_{XXC}$ itself. Introducing a local Hamiltonian on $W_{\rm emb}$ can therefore distinguish and couple these words, supplying a new dynamical structure that is invisible in the IS-shadow description. 
Importantly, $W_{\rm emb}$ is not one of the invariant subspaces defined by an IS. Rather, it provides a distinct decomposition of the Hilbert space that is transverse to the IS-based fragmentation. In particular, the product states spanning $W_{\rm emb}$ correspond to different frozen IS configurations. This distinction will be essential below: introducing nontrivial dynamics within $W_{\rm emb}$ can therefore connect different IS configurations while retaining $W_{\rm emb}$ itself as an invariant subspace. The IS word is not tied to fixed physical sites: the embedded basis contains full-length frozen words with different labels, and local exchanges between neighboring $0$ and $2$ states connect such words directly. Thus the first reduction brought by adding dynamics within $W_{\rm emb}$ changes the dynamical connectivity between the original fragmented sectors, rather than simply selecting one of them.

%%%%%%%%%%%%%%%%%%%%%%%%%%%%%%%%%%%%%%%%%%%%%%%%%%%%%%%
%%%%%%%%%%%%%%%%%%%%%%%%%%%%%%%%%%%%%%%%%%%%%%%%%%%%%%%
\section{Perturbed $XXC$ model from embedded spin-$1/2$ dynamics} 
%%%%%%%%%%%%%%%%%%%%%%%%%%%%%%%%%%%%%%%%%%%%%%%%%%%%%%%
%%%%%%%%%%%%%%%%%%%%%%%%%%%%%%%%%%%%%%%%%%%%%%%%%%%%%%%

The first reduction step consists of introducing couplings between originally disconnected IS-labeled sectors while retaining an exponentially large embedded invariant subspace. Within the general $XXC$ family, this is achieved by introducing dynamics inside the embedded invariant subspace $W_{\rm emb}$.
In the present work, we focus on the minimal spin-$1$ realization \eqref{eq:XXC}.

As shown in \eqref{eq:scalar}, the $XXC$ Hamiltonian acts as a scalar on $W_{\rm emb}$. One may then introduce the nontrivial dynamics within $W_{\rm emb}$ while preserving its invariance, thereby reducing the invariant-subspace structure by connecting different IS sectors.
The simplest realization of this idea is to introduce local perturbations that preserve $W_{\rm emb}$ and generate dynamics within it. 
We use the $XYZ$-type family as a concrete local realization, without asserting that a generic choice has an $\mathfrak{su}(2)$ decomposition into dynamically invariant irreducible sectors. A translationally invariant nearest-neighbor perturbation within $W_{\rm emb}$ includes spin-$1/2$ $XYZ$-type dynamics
\begin{align} \label{eq:emb-pert}
    &H_{\rm emb}(\widetilde{a},\widetilde{b},\widetilde{c},\widetilde{d}) = \sum_{j=1}^{L-1} (h_{\rm emb}(\widetilde{a},\widetilde{b},\widetilde{c},\widetilde{d}))_{j,j+1}, \\
    &h_{\rm emb}(\widetilde{a},\widetilde{b},\widetilde{c},\widetilde{d}) 
    = \widetilde{a} \sum_{s=0,2} |ss \rangle \langle ss| 
    + \widetilde{b} \sum_{\substack{s,t=0,2 \\s \neq t}} |st \rangle \langle st| \nonumber \\
    &\hspace{23mm}+ \widetilde{c} \sum_{\substack{s,t=0,2 \\s \neq t}} |st \rangle \langle ts| 
    + \widetilde{d} \sum_{\substack{s,t=0,2 \\s \neq t}} |ss \rangle \langle tt|, \nonumber
\end{align}
whose restricted dynamics within $W_{\rm emb}$ indeed is of spin-$1/2$ $XYZ$ type,
\begin{align} \label{eq:emb-rpert}
    H_{\rm emb}(\widetilde{a},\widetilde{b},\widetilde{c},\widetilde{d}) \big|_{W_{\rm emb}} 
    &= H_{XYZ}(\widetilde{a},\widetilde{b},\widetilde{c},\widetilde{d}) \\
    &= \sum_{j=1}^{L-1} (h_{XYZ}(\widetilde{a},\widetilde{b},\widetilde{c},\widetilde{d}))_{j,j+1}, 
\end{align}
where
\begin{align}
    h_{XYZ}(\widetilde{a},\widetilde{b},\widetilde{c},\widetilde{d}) &=  \widetilde{a} \sum_{s=\uparrow,\downarrow} |ss \rangle \langle ss| 
    + \widetilde{b} \sum_{\substack{s,t=\uparrow,\downarrow \\s \neq t}} |st \rangle \langle st| \nonumber \\
    &+ \widetilde{c} \sum_{\substack{s,t=\uparrow,\downarrow \\s \neq t}} |st \rangle \langle ts| 
    + \widetilde{d} \sum_{\substack{s,t=\uparrow,\downarrow \\s \neq t}} |ss \rangle \langle tt| \nonumber \\
    &= \frac{\widetilde{c}+\widetilde{d}}{2} \sigma^x \otimes \sigma^x
    + \frac{\widetilde{c}-\widetilde{d}}{2} \sigma^y \otimes \sigma^y \nonumber \\
    &+ \frac{\widetilde{a}-\widetilde{b}}{2} \sigma^z \otimes \sigma^z
    + \frac{\widetilde{a}+\widetilde{b}}{2} \bm{1}. 
\end{align}
Thus, preserving $W_{\rm emb}$ alone generally destroys the underlying IS fragmentation completely, leaving only the embedded spin-$1/2$ subspace $W_{\rm emb}$ invariant. The resulting dynamics therefore represents a reduction from exponentially many fragmented sectors to a single exponentially large invariant subspace. 

On the other hand, further imposing $U(1)$ conservation requires $\widetilde{d} = 0$ and reduces the dynamics within $W_{\rm emb}$ to the $XXZ$ type, thereby preserving its weight decomposition.
Remarkably, the $U(1)$-conserving restriction preserves not only the implicit spin-$1/2$ subspace $W_{\rm emb}$ but also the fully polarized IS subspaces $W_{\rm pol}^{(0)}$ and $W_{\rm pol}^{(2)}$ of the unperturbed $XXC$ model, corresponding to the IS labels $\lambda = \{0\}^{\ell_\lambda}$ and $\{2\}^{\ell_\lambda}$, respectively. 
Since the configuration exchange terms in \eqref{eq:emb-pert} nontrivially act only on adjacent $02$ and $20$ while they vanish on neighboring equal spins, the perturbation $H_{\rm emb}$ with $\widetilde{d} = 0$ does not violate the IS if it consists only of $0$s or $2$s. 
The perturbed Hamiltonian $H_{{\rm p}XXC}^{\rm pol} = H_{XXC} + H_{\rm emb}(\widetilde{a},\widetilde{b},\widetilde{c},0)$ therefore reproduces the perturbed $XXC$ model introduced in \cite{bib:KMPP2025}, in which $W_{\rm pol}^{(0,2)}$ remain invariant while the other IS subspaces are mixed.
The restricted dynamics within the polarized IS sectors remains of $XXZ$ type, including the diagonal contributions from the full local Hamiltonian. As demonstrated in \cite{bib:KMPP2025}, the dynamics in their complement is generically nonintegrable.

Thus, the previously studied weakly ergodicity-breaking perturbed $XXC$ model~\cite{bib:KMPP2025} emerges here as a consequence of preserving a $U(1)$-conserving spin-$1/2$ dynamics inside $W_{\rm emb}$.
Within the present nearest-neighbor spin-$1/2$ construction, the restricted dynamics is necessarily integrable, since a $U(1)$-conserving Hamiltonian of this form is of $XXZ$ type (Appendix~\ref{app:rigidity}). This integrability is therefore a consequence of the minimal embedded spin-$1/2$ structure rather than an additional assumption, and need not persist for more general embedded spaces.

When the embedded dynamics is compatible with an appropriate algebraic structure, its representation decomposition supplies candidates for the next reduction, independently of the original IS labels. In the $\mathfrak{su}(2)$ construction below, local mixing with the complement breaks the invariance of $W_{\rm emb}$ while annihilating a selected representation subspace. The construction introduced above has a direct interpretation from the viewpoint of invariant-subspace reduction. Since $W_{\rm emb}$ cuts across the IS-based fragmentation, the embedded dynamics generically connects states belonging to different IS configurations and thereby destroys the finer fragmented structure. At the same time, the perturbation preserves $W_{\rm emb}$, so that this exponentially large subspace remains invariant. The perturbed $XXC$ model can thus be viewed as a controlled reduction from IS fragmentation to a single exponentially large invariant subspace carrying nontrivial spin-$1/2$ dynamics.

%%%%%%%%%%%%%%%%%%%%%%%%%%%%%%%%%%%%%%%%%%%%%%%%%%%%%%%
%%%%%%%%%%%%%%%%%%%%%%%%%%%%%%%%%%%%%%%%%%%%%%%%%%%%%%%
\section{Emergent $XY$ scar model and rigidity of nonsymmetric $\mathfrak{su}(2)$-towers}
\label{sec:tower}
%%%%%%%%%%%%%%%%%%%%%%%%%%%%%%%%%%%%%%%%%%%%%%%%%%%%%%%
%%%%%%%%%%%%%%%%%%%%%%%%%%%%%%%%%%%%%%%%%%%%%%%%%%%%%%%

The embedded dynamics introduced in the previous section still preserves an exponentially large invariant subspace. 
The second reduction step consists of isolating a polynomial-dimensional invariant subspace from this larger invariant structure.

Within the general $XXC$ family, we achieve this reduction by introducing local couplings between $W_{\rm emb}$ and its complement while preserving a selected subspace of $W_{\rm emb}$.

The mixing construction is thus a separate step from choosing the dynamics inside $W_{\rm emb}$: a local realization of the latter does not guarantee a local leakage operator annihilating an arbitrarily chosen representation subspace. A natural choice for the minimal spin-$1$ realization is the $XXX$ branch of the embedded dynamics \eqref{eq:emb-pert}, \eqref{eq:emb-rpert}.
This branch retains the ordinary $\mathfrak{su}(2)$ representation structure within $W_{\rm emb}$ and therefore preserves the invariant subspaces generated from individual $\mathfrak{su}(2)$ highest-weight states. 
We now demonstrate that this representation-based invariant structure can be reduced by the coupling between $W_{\rm emb}$ and its complement further to a single $\mathfrak{su}(2)$ tower, thereby realizing a polynomial-dimensional invariant subspace.  
A more general discussion of the reduction hierarchy within the general $XXC$ family is presented in Appendix~\ref{app:general_xxc}. 

In order to realize such a reduction explicitly, let us consider the $U(1)$-conserving perturbation
\begin{align} 
    H_{\rm mix} = \sum_{j=1}^{L-1} (h_{\rm mix})_{j,j+1},
\end{align}
where
\begin{align} 
    h_{\rm mix} &= \big( \alpha_{02} |11 \rangle \langle 02| + \alpha^*_{02} |02 \rangle \langle 11| \big) \nonumber \\
    &+ \big( \alpha_{20} |11 \rangle \langle 20| + \alpha^*_{20} |20 \rangle \langle 11| \big). \label{eq:scar}
\end{align}
This perturbation couples $W_{\rm emb}$ to its complement and therefore destroys the embedded spin-$1/2$ invariant subspace in general. Nevertheless, $U(1)$ conservation guarantees that the two extremal fully polarized states $|\Uparrow \rangle :=|0 \rangle^{\otimes L}$ and $|\Downarrow \rangle := |2 \rangle^{\otimes L}$ remain exact eigenstates. 

We seek a perturbation that violates the invariance of $W_{\rm emb}$ while preserving the $\mathfrak{su}(2)$ tower generated from a fully polarized state, 
\begin{align}
    &W_{\rm max} = {\rm span}\{(T^-_{\rm tot})^n |\Uparrow \rangle \}_n 
    = {\rm span}\{(T^+_{\rm tot})^n |\Downarrow \rangle \}_n, \nonumber \\
    &T^{\pm}_{\rm tot} = \sum_{j=1}^L (S_j^{\pm})^2. \label{eq:ladder}
\end{align}
Since the fully symmetric $\mathfrak{su}(2)$ tower has support only in the local symmetric subspace ${\rm Sym}((\mathbb{C}^2)^{\otimes 2})$ on every neighboring pair, it is sufficient to require that $h_{\rm mix}$ annihilate the local symmetric representation of the effective spin-$1/2$ degrees of freedom in $W_{\rm emb}$. Denoting the corresponding projector by $P_{\rm sym}$, this condition reads 
\begin{align} \label{eq:sym_cancel}
    h_{\rm mix} P_{\rm sym} = 0,
\end{align}
with $P_{\rm sym} h_{\rm mix} = 0$ following from Hermiticity. 
For the perturbation in \eqref{eq:scar}, this condition reduces simply to
\begin{align} \label{eq:annihilation_cond}
    \alpha_{02} = -\alpha_{20} = \alpha.
\end{align}
Thus, this choice breaks the invariance of $W_{\rm emb}$ while preserving the fully symmetric $\mathfrak{su}(2)$ tower. Since the embedded $XXX$ Hamiltonian acts as a scalar on this tower and the mixing perturbation annihilates it, the tower remains degenerate. Its energy levels can be split by adding a term proportional to the Cartan generator. 

This annihilation mechanism is precisely the one realized in the spin-$1$ $XY$ scar model~\cite{bib:SI2019,bib:ISX2019,bib:OBCK2020,bib:M2026}. Its Hamiltonian can be decomposed as
\begin{align}
    H_{XY} = H_{\rm ann}+H_{\rm split},
\end{align}
where the annihilation term
\begin{align} \label{eq:XY_ann}
    &H_{\rm ann} = \sum_{j=1}^{L-1} (h_{\rm ann})_{j,j+1}, \nonumber \\ 
    &h_{\rm ann} = S_j^x S_{j+1}^x + S_j^y S_{j+1}^y
\end{align}
is related to the annihilation condition \eqref{eq:annihilation_cond} by the gauge transformation $V = \otimes_{j=1}^L e^{i\frac{\pi j}{2}S_j^z}$, while the energy splitting term is given by the $\mathfrak{su}(2)$ Cartan generator,
\begin{align} \label{eq:XY_split}
    H_{\rm split} = \frac{\mathcal{E}}{2} \sum_{j=1}^L S_j^z. 
\end{align}
On the fully symmetric tower $W_{\rm max}$, the untransformed Hamiltonian satisfies the rSGA
\begin{align} \label{eq:pol_rSGA}
    [H^{\rm pol}_{{\rm p}XXC} + H_{\rm mix} + H_{\rm split},\, T_{\rm tot}^{-}] + \mathcal{E}T_{\rm tot}^{-} 
    \big|_{W_{\rm max}}=0.
\end{align}
Under the gauge transformation $V$, this relation becomes
\begin{align} \label{eq:XY_rSGA}
    [H_{XY},\, T_{\rm tot}^{\pi;-}] + \mathcal{E} T_{\rm tot}^{\pi;-} \big|_{W^{(\pi)}_{\rm max}}=0,
\end{align}
where $W^{(\pi)}_{\rm max}: = VW_{\rm max}$ and $T_{\rm tot}^{\pi;-}=VT_{\rm tot}^{-}V^\dagger$~\cite{bib:SI2019,bib:OBCK2020,bib:M2026}.
In Appendix~\ref{app:XY}, we explicitly show how the local mixing perturbation is mapped to the annihilation term of the spin-$1$ $XY$ scar model.

This observation immediately raises a natural question: Is the fully symmetric HWS essential for isolating a single scar tower, or can the same reduction be performed for a tower built on a nonsymmetric HWS?

For a nonsymmetric HWS, the situation is qualitatively different. As we prove in Appendix~\ref{app:rigidity}, for even $L$, a nonmaximal-spin, nonsinglet $\mathfrak{su}(2)$ tower cannot be isolated from $W_{\rm emb}$ by a translationally invariant and $U(1)$-conserving nearest-neighbor perturbation. 
This obstruction depends on the imposed constraints. Indeed, nonmaximal-spin $\mathfrak{su}(2)$ scar towers have been constructed in translationally invariant spin-$1$ chains with next-nearest-neighbor interactions and terms that break $U(1)$ conservation~\cite{bib:OBCK2020}. 

% Figure here
\begin{figure*}[t]
\centering
\begin{tikzpicture}[
    >=stealth,
    box/.style={
        draw,
        rounded corners=3pt,
        align=center,
        minimum height=0.9cm,
        text width=5.2cm,
        inner sep=4pt
    },
    topbox/.style={
        draw,
        rounded corners=3pt,
        align=center,
        minimum height=0.9cm,
        text width=7.2cm,
        inner sep=4pt
    },
    arrow/.style={
        ->,
        line width=0.7pt
    },
    node distance=0.9cm and -2.5cm
]

%------------------------------------------------
% Starting condition
%------------------------------------------------
\node[topbox] (tower) {
    \textbf{Preserve a single $\mathfrak{su}(2)$ tower}\\
    {\small translational invariance \\+ $U(1)$ conservation \\
    + nearest-neighbor locality}
};

%------------------------------------------------
% Two classes of HWS
%------------------------------------------------
\node[box, below left=of tower] (sym) {
    \textbf{Fully symmetric HWS}\\
    maximal spin
};

\node[box, below right=of tower] (nonsym) {
    \textbf{Nonsymmetric HWS}\\
    nonmaximal spin, nonsinglet
};

%------------------------------------------------
% Consequences
%------------------------------------------------
\node[box, below=of sym] (break) {
    $W_{\rm emb}$ \textbf{can be broken}\\
    while the tower survives
};

\node[box, below=of nonsym] (restore) {
    $W_{\rm emb}$ \textbf{is restored}\\
    leakage is forbidden
};

%------------------------------------------------
% Final structures
%------------------------------------------------
\node[below=0.7cm of break, align=center] (scar) {
    \textbf{isolated scar tower}
};

\node[below=0.7cm of restore, align=center] (extended) {
    \textbf{extended invariant}\\
    \textbf{spin-$1/2$ space}
};

%------------------------------------------------
% Arrows
%------------------------------------------------
\draw[arrow] (tower.south) -- ++(0,-0.3)
    -| (sym.north);

\draw[arrow] (tower.south) -- ++(0,-0.3)
    -| (nonsym.north);

\draw[arrow] (sym) -- (break);
\draw[arrow] (nonsym) -- (restore);

\draw[arrow] (break) -- (scar);
\draw[arrow] (restore) -- (extended);

\end{tikzpicture}

\caption{
Schematic illustration of the rigidity of the invariant-subspace
reduction.
A fully symmetric tower can remain invariant after $W_{\rm emb}$ is
broken, whereas preserving a nonmaximal-spin, nonsinglet
$\mathfrak{su}(2)$ tower under the stated local constraints restores
the entire embedded spin-$1/2$ space.
}
\label{fig:rigidity}
\end{figure*}
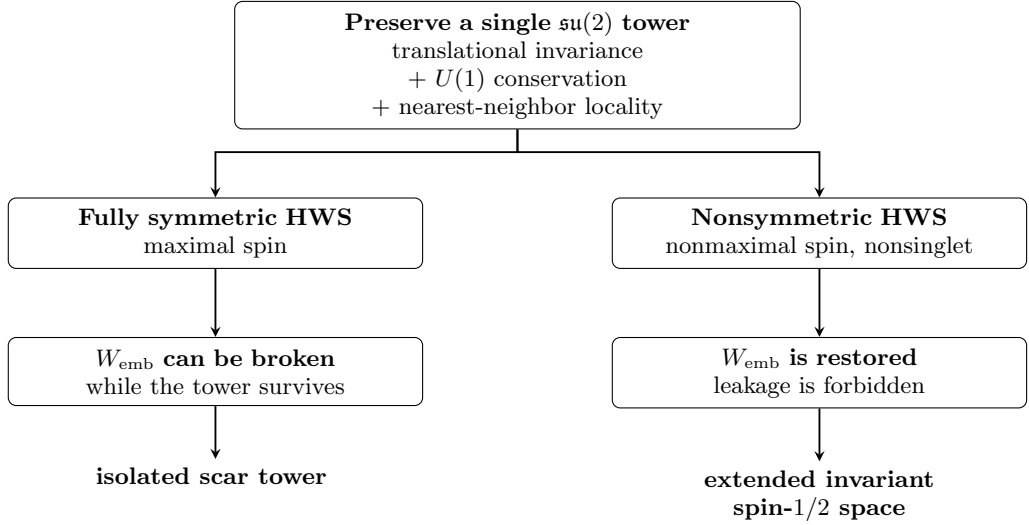

Within $W_{\rm emb}$, the restored $U(1)$-conserving dynamics is of $XXZ$ type, possibly with a DM phase that can be absorbed into the corresponding twisted $\mathfrak{su}(2)$ generators by a local gauge transformation. 

Thus, within the translationally invariant, $U(1)$-conserving nearest-neighbor setting considered here, the fully symmetric maximal-spin tower is exceptional: it can remain intact after $W_{\rm emb}$ itself is broken, whereas a generic nonsymmetric, nonmaximal-spin, nonsinglet $\mathfrak{su}(2)$ tower cannot be isolated under the same assumptions. 

The hierarchy discussed so far can therefore be summarized as successive reductions of the same representation-based invariant structure. The fragmented $XXC$ model, the perturbed $XXC$ model with an embedded spin-$1/2$ invariant space, and the spin-$1$ $XY$ scar model are connected through this reduction, terminating in a single exceptional fully symmetric scar tower.

%%%%%%%%%%%%%%%%%%%%%%%%%%%%%%%%%%%%%%%%%%%%%%%%%%%%%%%
%%%%%%%%%%%%%%%%%%%%%%%%%%%%%%%%%%%%%%%%%%%%%%%%%%%%%%%
\section{From polarized to N\'{e}el}
\label{sec:pol_alt}
%%%%%%%%%%%%%%%%%%%%%%%%%%%%%%%%%%%%%%%%%%%%%%%%%%%%%%%

The intermediate invariant subspace $W_{\rm emb}$ plays a central role in relating seemingly different constructions of invariant subspaces.
Since the dynamics within $W_{\rm emb}$ can be expressed in different bases, the same invariant-subspace structure may appear in different forms when expressed in terms of the physical spin configurations.
This observation allows us to relate constructions based on different reference states through unitary transformations within $W_{\rm emb}$.
The correspondence between the two constructions is based on a unitary transformation that preserves the unperturbed $XXC$ Hamiltonian. As shown below, the transformation within $W_{\rm emb}$ can be extended to a unitary operator on the full Hilbert space that commutes with $H_{XXC}$. This does not, however, imply unitary equivalence between the full perturbed Hamiltonians.

The $U(1)$-conserving embedded perturbation preserves the polarized IS sectors, whereas the pair-flipping perturbation preserves the alternating IS sectors~\cite{bib:KMPP2025}. The latter is not $U(1)$ conserving in the original spin-$1$ basis, while its dynamics in the preserved sectors is related to the former by a staggered transformation.

For even $L$, define the spin flips within $W_{\rm emb}$ by
\begin{align} \label{eq:staggered_U}
    U^{(1)}
    &= \bigotimes_{j=1}^{L/2}\sigma_{2j-1}^{x}, \qquad
    U^{(2)}
    = \bigotimes_{j=1}^{L/2}\sigma_{2j}^{x}.
\end{align}
Each nearest-neighbor bond contains exactly one flipped effective spin. Consequently, the exchange and pair-flipping terms are interchanged, while the two diagonal coefficients are exchanged. 
The resulting pair-flipping perturbed $XXC$ Hamiltonian is
\begin{align} \label{eq:alt_pXXC}
    H_{{\rm p}XXC}^{\rm alt}
    = H_{XXC} + H_{\rm emb}(\widetilde{a},\widetilde{b},0,\widetilde{d}).
\end{align}
The embedded perturbations satisfy
\begin{align} \label{eq:c_to_d}
    H_{\rm emb}(\widetilde{a},\widetilde{b},0,\widetilde{d})
    \big|_{W_{\rm emb}}
    &= U^{(\nu)}
    H_{XXZ}(\widetilde{b},\widetilde{a},\widetilde{d})
    (U^{(\nu)})^\dagger \nonumber \\
    &= U^{(\nu)}
    H_{\rm emb}(\widetilde{b},\widetilde{a},\widetilde{d},0)
    \big|_{W_{\rm emb}}
    (U^{(\nu)})^\dagger,
\end{align}
where $\nu=1,2$.
The scalar action of $H_{XXC}$ on $W_{\rm emb}$~\eqref{eq:scalar} is unchanged. The pair-flipping perturbation in \eqref{eq:alt_pXXC} coincides with the one previously constructed from the IS-based perspective~\cite{bib:KMPP2025}. 

Thus, the polarized and alternating perturbed $XXC$ models, originally obtained by preserving different families of IS sectors, are related by a unitary transformation of their dynamics within the common invariant subspace $W_{\rm emb}$. This equivalence of the restricted dynamics does not imply unitary equivalence between the full perturbed spin-$1$ Hamiltonians.

The two fully polarized states $|\Uparrow\rangle$ and $|\Downarrow\rangle$ are mapped to the two N\'{e}el configurations,
\begin{align} \label{eq:neel_states}
    |\text{N\'{e}el}\,1\rangle
    &= U^{(1)}|\Uparrow\rangle
    = U^{(2)}|\Downarrow\rangle, \nonumber \\
    |\text{N\'{e}el}\,2\rangle
    &= U^{(2)}|\Uparrow\rangle
    = U^{(1)}|\Downarrow\rangle.
\end{align}
One concrete realization of the staggered correspondence is a bond-dimension-two matrix-product operator (MPO) tracking the order of the $0/2$ species; see Appendix~\ref{app:restricted_correspondence}.

Thus, the polarized and alternating configurations serve as extremal vectors in two local realizations of the same embedded representation. In particular, the correspondence concerns the invariant sectors and their restricted dynamics, not an equivalence of the complementary Hilbert-space dynamics.

Although the staggered spin flips $U^{(\nu)}$ preserve the locality and interaction range of operators acting within $W_{\rm emb}$, a site-dependent transformation does not generally preserve one-site translational invariance. For the nearest-neighbor density in \eqref{eq:emb-pert}, the two possible choices of the flipped site on each bond yield the same transformed density. Consequently, the transformed perturbation in \eqref{eq:c_to_d} remains one-site translationally invariant. The condition for such a local realization, and its generalization to other patterns of local spin flips, are discussed in Appendix~\ref{app:local_realization}.

The staggered transformation can also be applied to the selected scar tower. As shown in the next section, transforming the local leakage amplitudes on their $W_{\rm emb}$ domain produces a nearest-neighbor mixing perturbation annihilating the N\'{e}el tower. No full-space intertwiner of the $XXC$ Hamiltonian is needed for this construction.

%%%%%%%%%%%%%%%%%%%%%%%%%%%%%%%%%%%%%%%%%%%%%%%%%%%%%%%
%%%%%%%%%%%%%%%%%%%%%%%%%%%%%%%%%%%%%%%%%%%%%%%%%%%%%%%
\section{Scar towers on N\'{e}el references}
\label{sec:neel_scar}
%%%%%%%%%%%%%%%%%%%%%%%%%%%%%%%%%%%%%%%%%%%%%%%%%%%%%%%
%%%%%%%%%%%%%%%%%%%%%%%%%%%%%%%%%%%%%%%%%%%%%%%%%%%%%%%

We now carry the staggered transformation through the second reduction, from the embedded spin-$1/2$ space to a single scar tower. This step requires only a unitary transformation within $W_{\rm emb}$, given by $U^{(\nu)}$, and does not require a unitary transformation of the full Hilbert space that commutes with $H_{XXC}$.

We write the leakage part of the local mixing term \eqref{eq:scar}, with the annihilation condition~\eqref{eq:annihilation_cond}, as
\begin{align} \label{eq:pol_leakage}
    (h_{w \to w^{\bot}})_{j,j+1}
    &= \alpha \big( |11\rangle (\langle 02| - \langle 20|) \big)_{j,j+1}, 
\end{align}
which maps $W_{\rm emb}$ to its complement. 
For each $\nu=1,2$, we define the transformed local leakage operator by its action on $W_{\rm emb}$:
\begin{align} \label{eq:transformed_leakage}
    (h'_{w \to w^{\bot}})_{j,j+1}
    &= (h_{w \to w^{\bot}})_{j,j+1} ( U_j^{(\nu)} \otimes U_{j+1}^{(\nu)} )^\dagger,
\end{align}
which again maps $W_{\rm emb}$ to its complement. Here, $U_j^{(\nu)} = \sigma_j^x$ if $j = \nu \pmod 2$ and $U_j^{(\nu)} = \bm{1}$ otherwise. 
Since the original local leakage operator \eqref{eq:pol_leakage} has support on the two-site embedded space, the transformed leakage operator \eqref{eq:transformed_leakage} defines an ordinary nearest-neighbor operator on the spin-$1$ chain. Its definition does not depend on how $U^{(\nu)}$ might be extended outside $W_{\rm emb}$.

For $\nu = 1$, the leakage operators \eqref{eq:pol_leakage} on successive bonds are transformed as
\begin{align} \label{eq:neel_leakage}
    (h'_{w \to w^{\bot}})_{2j-1,2j}^{(1)}
    &= \alpha \big( |11 \rangle (\langle 22| - \langle 00|) \big)_{2j-1,2j}, \nonumber \\
    (h'_{w \to w^{\bot}})_{2j,2j+1}^{(1)}
    &= \alpha \big( |11 \rangle (\langle 00| - \langle 22|) \big)_{2j,2j+1}.
\end{align}
For the $\nu = 2$ transformation, we observe the reversed signed leakage terms. Thus the transformed mixing is nearest-neighbor and two-site periodic; one-site translational invariance is not required. 

Thus, the second reduction is achieved by adding the transformed mixing perturbation
\begin{align} \label{eq:neel_mix}
    H'^{(\nu)}_{\rm mix}
    &= \sum_{j=1}^{L-1} (h'^{(\nu)}_{w \to w^{\bot}})_{j,j+1}
    + (h'^{(\nu)}_{w \to w^{\bot}})^\dagger_{j,j+1})
\end{align}
to the alternating perturbed $XXC$ Hamiltonian \eqref{eq:alt_pXXC}. 

The annihilation condition can be established in parallel with the polarized construction. 
On each bond, define the transformed local symmetric projector by
\begin{align}
    P'^{(\nu)}_{\rm sym}
    &= ( U_j^{(\nu)} \otimes U_{j+1}^{(\nu)} )
    P_{\rm sym}
    ( U_j^{(\nu)} \otimes U_{j+1}^{(\nu)} )^\dagger.
\end{align}
Since the original leakage satisfies $(h_{W \to W^{\bot}})_{j,j+1} P_{\rm sym}=0$, the transformed leakage obeys
\begin{align}
    (h'^{(\nu)}_{w \to w^{\bot}})_{j,j+1} P_{\rm sym}^{\prime(\nu)}
    &= (h_{w \to w^{\bot}})_{j,j+1}P_{\rm sym} ( U_j^{(\nu)} \otimes U_{j+1}^{(\nu)} )^\dagger \nonumber \\
    &= 0.
\end{align}
Together with the Hermitian-conjugate contribution, this implies
\begin{align} \label{eq:annihilation_cond}
    h'^{(\nu)}_{\rm mix} P'^{(\nu)}_{\rm sym} = 0.
\end{align}
Thus, the transformed mixing perturbation annihilates the tower subspace $W'^{(\nu)}_{\rm max} = U^{(\nu)} W_{\rm max}$ spanned by the tower states on a N\'{e}el reference state \eqref{eq:neel_states} 
\begin{align} \label{eq:neel_tower}
    |\Psi_n^{\prime(\nu)}\rangle
    &= (T_{\rm tot}^{\prime-\,(\nu)})^n
    |\text{N\'{e}el}\,\nu\rangle, 
\end{align}
which are constructed by the staggered ladder operator $T_{\rm tot}^{\prime-\,(\nu)} = U^{(\nu)}T_{\rm tot}^- (U^{(\nu)})^\dagger$. 
Since the staggeredly transformed embedded $XXX$ Hamiltonian acts as a scalar on $W_{\rm max}^{\prime(\nu)}$, the full Hamiltonian preserves this tower.

An equally spaced tower spectrum is obtained by adding a term proportional to the staggered Cartan generator 
\begin{align} \label{eq:neel_cartan}
    T'^{z\,(\nu)}_{\rm tot}
    &= U^{(\nu)} T_{\rm tot}^z (U^{(\nu)})^\dagger
    = \sum_{j=1}^L (-1)^{j+\nu-1} S_j^z.
\end{align}
The rSGA structure of the untransformed model \eqref{eq:pol_rSGA} is inherited by the N\'{e}el-reference tower through the unitary equivalence of the restricted embedded dynamics and the annihilation of the transformed tower by the local leakage operators~\eqref{eq:annihilation_cond}. 

The N\'{e}el construction therefore illustrates the constructive aspect of the hierarchy: a unitary relation within an exponentially large embedded invariant space can be carried to a polynomial-dimensional invariant tower by transforming the local leakage on its embedded domain. The resulting full Hamiltonians need not be unitarily equivalent.

%%%%%%%%%%%%%%%%%%%%%%%%%%%%%%%%%%%%%%%%%%%%%%%%%%%%%%%
%%%%%%%%%%%%%%%%%%%%%%%%%%%%%%%%%%%%%%%%%%%%%%%%%%%%%%%
\section{Conclusion}
\label{sec:conclusion}
%%%%%%%%%%%%%%%%%%%%%%%%%%%%%%%%%%%%%%%%%%%%%%%%%%%%%%%
%%%%%%%%%%%%%%%%%%%%%%%%%%%%%%%%%%%%%%%%%%%%%%%%%%%%%%%

We have developed a hierarchy in which local deformations reorganize invariant-subspace structures. In the spin-$1$ $XXC$ chain, the embedded spin-$1/2$ space $W_{\rm emb}$ contains all full-length irreducible-string configurations. Local dynamics introduced within this space connects configurations with different IS labels and organizes them according to the dynamics of the embedded spin-$1/2$ degrees of freedom. A subsequent local mixing with its complement can break the invariance of $W_{\rm emb}$ while retaining a selected polynomial-dimensional $\mathfrak{su}(2)$ tower. This connects the fragmented $XXC$ model, partially solvable perturbed $XXC$ models, and the spin-$1$ $XY$ scar model through two successive reductions.

The hierarchy has both a restrictive and a constructive consequence. Its rigidity is expressed, for even $L$, by the requirement that preserving a nonmaximal-spin nonsymmetric $\mathfrak{su}(2)$ tower under translational invariance, $U(1)$ conservation, and nearest-neighbor locality also preserves the larger embedded spin-$1/2$ space. Conversely, a unitary transformation within $W_{\rm emb}$ can be transferred to the subsequent reduction by transforming the embedded dynamics and the action of the local leakage operators on $W_{\rm emb}$. We implemented this correspondence for polarized and alternating reference configurations and constructed an exact N\'{e}el-reference scar tower. The correspondence concerns the restricted dynamics and the selected invariant subspaces, without assuming unitary equivalence of the full Hamiltonians.

These two results characterize complementary aspects of representation-based reduction. Rigidity constrains which levels can be separated under local symmetry requirements, while constructive correspondence shows how the freedom of representation realization in an intermediate invariant space can generate different lower-level invariant structures. The N\'{e}el tower is an explicit realization of this mechanism, not its defining principle.

The hierarchy also admits a direct dynamical interpretation: successive reductions progressively restrict the class of initial states that can access nonthermal dynamics supported by the invariant subspaces.
Extending this perspective to medium-range interactions, higher-spin embedded local spaces, and other algebraically organized invariant sectors may reveal richer hierarchies and clarify how their structure controls initial-state-dependent nonthermal dynamics. 
It would also be interesting to classify site-dependent realizations of equivalent restricted dynamics, even when their full many-body spectra differ.

\acknowledgements

This work was supported by JSPS KAKENHI Grants No. JP26K06937.

%%%%%%%%%%%%%%%%%%%%%%%%%%%%%%%%%%%%%%%%%%%%%%%%%%%%%%%
%  starting appendix
%%%%%%%%%%%%%%%%%%%%%%%%%%%%%%%%%%%%%%%%%%%%%%%%%%%%%%%
\appendix
%%%%%%%%%%%%%%%%%%%%%%%%%%%%%%%%%%%%%%%%%%%%%%%%%%%%%%%
%%%%%%%%%%%%%%%%%%%%%%%%%%%%%%%%%%%%%%%%%%%%%%%%%%%%%%%
\section{Rigidity of nonsymmetric $\mathfrak{su}(2)$ towers}
\label{app:rigidity}
%%%%%%%%%%%%%%%%%%%%%%%%%%%%%%%%%%%%%%%%%%%%%%%%%%%%%%%
%%%%%%%%%%%%%%%%%%%%%%%%%%%%%%%%%%%%%%%%%%%%%%%%%%%%%%%

In this appendix, we establish, for even $L$, a rigidity property of a nonsymmetric $\mathfrak{su}(2)$ tower embedded in $W_{\rm emb} = {\rm span}\{|0 \rangle, |2 \rangle\}^{\otimes L}$. 
We first prove the statement for the ordinary $\mathfrak{su}(2)$ tower and subsequently describe its consequences for gauge-conjugated Hamiltonians. We show for even $L$ that a translationally invariant and $U(1)$-conserving nearest-neighbor perturbation preserving a tower built on a nonmaximal-spin, nonsinglet HWS cannot couple $W_{\rm emb}$ to its complement. Consequently, the entire embedded spin-$1/2$ subspace remains invariant. The $U(1)$-conserving nearest-neighbor dynamics restored within this subspace is necessarily of spin-$1/2$ $XXZ$ type, supplemented by magnetic-field, Dzyaloshinskii–Moriya (DM), and boundary terms, and is therefore integrable under open boundary conditions. 
Finally, we describe the gauge-covariant form of this result for Hamiltonians obtained by conjugating the class considered above.

We consider a perturbation that is allowed to couple $W_{\rm emb}$ to its complement, while requiring a selected $\mathfrak{su}(2)$ tower
\begin{align}
    &W_{\Omega} = {\rm span}\{ |\Psi^{\Omega}_n \rangle \}_n, \nonumber \\
    &|\Psi^{\Omega}_n \rangle
    = (T_{\rm tot}^-)^n |{\rm HWS}_{\Omega} \rangle
\end{align}
to remain invariant. Since the HWS $|{\rm HWS}_{\Omega} \rangle$ belongs to neither maximal-spin nor singlet, the tower subspace $W_{\Omega}$ satisfies $2 \leq {\rm dim}W_{\Omega} \leq L$.

We first prove that preserving the tower $\{|\Psi^{\Omega}_n\rangle\}_n$ forces the entire subspace $W_{\rm emb}$ to remain invariant. 
For the proof, it is useful to introduce the block decomposition of the 
local perturbation with respect to the local subspace $w = {\rm span}\{|0 \rangle, |2 \rangle\}^{\otimes 2}$ and its orthogonal complement $w^{\bot}$. The global embedded subspace $W_{\rm emb}$ consists precisely of the many-body states whose local support on every bond is contained in $w$. 

We decompose the local perturbation as 
\begin{align} \label{eq:decomp_h}
    h_{\rm pert}
    = h_w + h_{{w^{\bot}}}
    + h_{w^{\bot} \to w} + h_{w \to w^{\bot}}.
\end{align}
Here, $h_w$ and $h_{w^\bot}$ act within $w$ and $w^\bot$, respectively, while $h_{w^\bot\to w}$ and $h_{w\to w^\bot}$ denote the off-diagonal blocks. By Hermiticity, $h_{w^{\bot} \to w}^{\dag} = h_{w \to w^{\bot}}$, and therefore we only discuss $h_{w \to w^{\bot}}$. The mixing term introduced in \eqref{eq:scar} consists only of the off-diagonal blocks, $h_{\rm mix} = h_{w\to w^\perp} + h_{w^\perp\to w}$. 

By $U(1)$ conservation, the leakage term is restricted to
\begin{align}
     h_{w \to w^{\bot}}
     &= \alpha_{02} |11 \rangle \langle 02|
     + \alpha_{20} |11 \rangle \langle 20| \nonumber \\
     &= c_+ |11 \rangle \langle \psi_+|
     + c_- |11 \rangle \langle \psi_-|,
\end{align}
where $c_{\pm} = (\alpha_{02} \pm \alpha_{20})/\sqrt{2}$ and $|\psi_{\pm} \rangle = (|02 \rangle \pm |20 \rangle)/\sqrt{2}$. 
For each bond $(j,j+1)$, the image of $W_{\rm emb}$ under $(h_{w \to w^{\bot}})_{j,j+1}$ lies in the subspace $\mathcal K_j$ spanned by basis configurations with $1$ precisely at sites $j,j+1$ and $0$ or $2$ at every other site. Since $\mathcal K_j \perp \mathcal K_k$ for $j\neq k$, leakage contributions from different bonds cannot cancel, even for superpositions of embedded basis states. Hence, invariance of the tower requires the action of $h_{w \to w^{\bot}}$ to vanish separately on every bond. 

Since the selected tower is a nonsinglet $\mathfrak{su}(2)$ representation, its HWS has positive magnetization. For even $L$, the number of $0$s in every basis configuration with nonzero amplitude exceeds $L/2$, and hence at least one bond $(j,j+1)$ must contain adjacent $0$s under open boundary conditions: 
\begin{align} \label{eq:00}
    (|00 \rangle \langle 00|)_{j,j+1} |\Psi_0^\Omega \rangle
    \neq 0.
\end{align}
Bondwise annihilation of the HWS requires
\begin{align}
    (h_{w \to w^\bot})_{j,j+1} |\Psi_0^\Omega \rangle
    = 0,
\end{align}
while bondwise annihilation of the first descendant gives
\begin{align}
    &(h_{w \to w^\bot})_{j,j+1}
    \big((S_j^-)^2 + (S_{j+1}^-)^2\big)
    |\Psi_0^\Omega \rangle \nonumber \\
    &= 2\sqrt{2} c_+
    (|11 \rangle \langle 00|)_{j,j+1} |\Psi_0^\Omega \rangle \nonumber \\
    &= 0.
\end{align}
Since the HWS contains at least one bond with adjacent $0$s \eqref{eq:00}, this condition can hold on every bond only when 
\begin{align}
    c_+ = 0.
\end{align}

The remaining leakage term is therefore
\begin{align}
    h_{w \to w^\bot}
    = c_- |11 \rangle \langle \psi_-|.
\end{align}
Bondwise annihilation of the HWS then requires
\begin{align}
    &(h_{w \to w^\bot})_{j,j+1} |\Psi_0^\Omega \rangle \nonumber \\
    &=c_- (|11 \rangle \langle \psi_-|)_{j,j+1}
    |\Psi_0^\Omega \rangle \nonumber \\
    &= 0.
\end{align}
If $c_- \neq 0$, this condition can be satisfied on every bond only when
\begin{align}
    (\langle \psi_-|)_{j,j+1} |\Psi_0^\Omega \rangle = 0,
    \qquad \forall j.
\end{align}
Thus, the HWS has no antisymmetric component on any neighboring pair, forcing every neighboring pair to belong to the symmetric triplet subspace. Equivalently, $|\Psi_0^\Omega \rangle$ is invariant under every adjacent transposition. Thus, the HWS must be fully symmetric and therefore belong to the maximal-spin representation. This contradicts the assumption that the selected tower is nonmaximal-spin, and hence we have 
\begin{align}
    c_- = 0.
\end{align}

Consequently, preserving a nonsinglet, nonmaximal-spin tower requires $c_+ = c_- = 0$, or equivalently, 
\begin{align}
    \alpha_{02} = \alpha_{20} = 0.
\end{align}
Thus, invariance of a nonsinglet, nonmaximal-spin tower prohibits mixing between $W_{\rm emb}$ and $W_{\rm emb}^{\bot}$, and the entire embedded spin-$1/2$ subspace $W_{\rm emb}$ necessarily remains invariant.

We next characterize the dynamics restored within $W_{\rm emb}$. After imposing the tower-preservation condition, the perturbation \eqref{eq:decomp_h} has the block-diagonal structure 
\begin{align} \label{eq:diagonal_mix}
    h_{\rm pert} = h_w + h_{{w^{\bot}}}.
\end{align}
Consequently, the Hamiltonian of the full perturbed model restricted to $W_{\rm emb}$ is written as 
\begin{align}
    &\big(H^{\rm pol}_{{\rm p}XXC} + H_{\rm pert}\big) \big|_{W_{\rm emb}} \nonumber \\
    &= a(L-1) \bm{1} + \sum_{j=1}^{L-1} (h_{XXX} + h_w)_{j,j+1},
\end{align}
where $H_{\rm pert} = \sum_{j=1}^{L-1} (h_{\rm pert})_{j,j+1}$. 

Due to $U(1)$ invariance, the general Hermitian two-site density of the restricted Hamiltonian can be written as 
\begin{align} \label{eq:general_spinhalf}
    h_{XXX} + h_w
    &= h_{XXZ}(a,b,c)
    + h_{\rm mag}(m) \nonumber \\
    &\quad + h_{\rm DM}(D) + h_{\rm tel}(J_{\rm tel})
    + \text{const},
\end{align}
where $h_{XXZ}(a,b,c)$ is given in \eqref{eq:h_XXZ}, and
\begin{align}
    &h_{\rm mag}(m)
    = \frac{m}{2} (\sigma^z \otimes \bm{1} + \bm{1} \otimes \sigma^z), \nonumber \\
    &h_{\rm DM}(D)
    = D (\sigma^x \otimes \sigma^y - \sigma^y \otimes \sigma^x), \nonumber \\
    &h_{\rm tel}(J_{\rm tel})
    = J_{\rm tel} (\sigma^z \otimes \bm{1} - \bm{1} \otimes \sigma^z). 
\end{align}
This is the local density of the spin-$1/2$ $XXZ$ chain supplemented by magnetic-field, Dzyaloshinskii–Moriya (DM), and telescoping terms. Under open boundary conditions, the telescoping term produces longitudinal boundary magnetic fields, while a uniform DM phase can be removed by a site-dependent gauge transformation without introducing a boundary twist. The resulting Hamiltonian is an open $XXZ$ chain with a uniform bulk magnetic field and longitudinal boundary fields, which is integrable for arbitrary values of the boundary fields~\cite{bib:S1988,bib:DG1994}. Thus, preservation of the nonsymmetric tower necessarily restores $W_{\rm emb}$ together with integrable spin-$1/2$ dynamics therein.

We finally show that the rigidity obtained above does not depend on the choice of unitary gauge realization of the $\mathfrak{su}(2)$ tower. Consider a local gauge transformation within the embedded spin-$1/2$ space $W_{\rm emb}$, 
\begin{align}
    G_{\bm{\theta}}
    = \otimes_{j=1}^L e^{i\theta_j \sigma_j^z/2}. 
\end{align}
The embedded spin-$1/2$ subspace $W_{\rm emb}$ is preserved by this gauge transformation, while the selected $\mathfrak{su}(2)$ tower states are twisted as 
\begin{align}
    &|\Psi_n^{\Omega}(\bm{\theta}) \rangle 
    = (T_{\rm tot}^-(\bm{\theta}))^n |{\rm HWS}_{\Omega}(\bm{\theta}) \rangle, \nonumber \\
    &T_{\rm tot}^-(\bm{\theta})
    = G_{\bm{\theta}} T^-_{\rm tot}
    G_{\bm{\theta}}^\dagger = \sum_{j=1}^L e^{-i\theta_j}(S_j^-)^2,  
\end{align}
which form the twisted tower subspace $W_{\Omega}^{(\bm{\theta})} = G_{\bm{\theta}} W_{\Omega}$. 
Consequently, the twisted tower subspace $W_{\Omega}^{(\bm{\theta})}$ carries the corresponding unitarily conjugated $\mathfrak{su}(2)$ representation. In particular, a nonmaximal-spin, nonsinglet irreducible representation is mapped to the corresponding nonmaximal-spin, nonsinglet representation of the twisted $\mathfrak{su}(2)$ algebra.

Let $H_0$ belong to the translationally invariant, $U(1)$-conserving nearest-neighbor class covered by the preceding proof, and suppose that it preserves $W_{\Omega}$. For the gauge-conjugated Hamiltonian $H_{\bm{\theta}} = G_{\bm{\theta}} H_0 G_{\bm{\theta}}^\dagger$, the tower $W_{\Omega}^{(\bm{\theta})}$ is preserved. The rigidity result for $H_0$ implies that $W_{\rm emb}$ is invariant under $H_0$, and therefore also under $H_{\bm{\theta}}$, since $G_{\bm{\theta}} W_{\rm emb}=W_{\rm emb}$.

The gauge transformation may produce bond-dependent transverse and DM couplings, while a linear gauge $\theta_j = j\phi$ preserves the uniformity of the local density. In either case, the restricted dynamics remains gauge equivalent to an open $XXZ$ chain with longitudinal magnetic fields and is therefore integrable. Thus the larger invariant space and its restricted integrable dynamics are preserved along the gauge orbit of the Hamiltonians covered by the rigidity statement.

%%%%%%%%%%%%%%%%%%%%%%%%%%%%%%%%%%%%%%%%%%%%%%%%%%%%%%%
%%%%%%%%%%%%%%%%%%%%%%%%%%%%%%%%%%%%%%%%%%%%%%%%%%%%%%%
\section{Restricted staggered correspondence}
\label{app:restricted_correspondence}
%%%%%%%%%%%%%%%%%%%%%%%%%%%%%%%%%%%%%%%%%%%%%%%%%%%%%%%

The construction in Secs.~\ref{sec:pol_alt} and~\ref{sec:neel_scar} is formulated entirely in terms of the embedded subspace $W_{\rm emb}$ and its local coupling to the complement. Here we provide explicit operator realizations of the correspondence, including a full-space unitary transformation between IS sectors and the local transformations of the embedded dynamics and leakage.

We first construct a site-independent matrix-product operator (MPO) that tracks the order of the $0/2$ species. The polarized-to-alternating correspondence is a particular example of this construction.
Let $|U\rangle,|D\rangle$ span a two-dimensional auxiliary space $\mathbb{C}^2$. Using the operators defined by $P_j = (|0 \rangle \langle 0| + |2 \rangle \langle 2|)_j$ and $X_j = (|0 \rangle \langle 2| + |2 \rangle \langle 0|)_j$, we define the local tensor as 
\begin{align} \label{eq:staggered_mpo_tensor}
    \mathcal M_j
    &= |U \rangle \langle D|\otimes P_j
    + |D \rangle \langle U|\otimes X_j
    + \bm{1} \otimes |1 \rangle_j \langle 1|.
\end{align}
For different choices of auxiliary boundary conditions, we have four different operators 
\begin{align}
    \mathcal U_{ab}
    &= \langle a|\mathcal M_1 \cdots \mathcal M_L|b \rangle,
    \qquad a,b\in\{U,D\}.
\end{align}
The auxiliary state changes whenever a $0/2$ species is encountered and remains unchanged at a site occupied by $1$. Thus, the tensor tracks the position in the ordered IS word rather than the physical-site parity. On $W_{\rm emb}$, the order of the $0/2$ species coincides with the physical-site order.

For even $L$, the two diagonal contractions reproduce the staggered flips on the entire embedded space $W_{\rm emb}$:
\begin{align} \label{eq:staggered_mpo_identity}
    &\mathcal U_{DD}\big|_{W_{\rm emb}}
    = \bigotimes_{j\,{\rm odd}} X_j
       \bigotimes_{j\,{\rm even}} P_j
    = U^{(1)}, \nonumber \\
    &\mathcal U_{UU} \big|_{W_{\rm emb}}
    = \bigotimes_{j\,{\rm odd}} P_j
       \bigotimes_{j\,{\rm even}} X_j
    = U^{(2)}. 
\end{align}
For configurations with an odd number of $0/2$ species, the nonzero contractions instead have off-diagonal auxiliary boundary indices.

The same tensor also gives unitary transformations on the full Hilbert space by summing over the left auxiliary boundary state:
\begin{align} \label{eq:full_is_intertwiner}
    \widetilde U^{(1)}
    &= (\langle U| + \langle D|)
    \mathcal M_1 \cdots \mathcal M_L|D \rangle, \\
    \widetilde U^{(2)}
    &= (\langle U| + \langle D|)
    \mathcal M_1 \cdots \mathcal M_L|U \rangle.
\end{align}
For each computational-basis configuration, exactly one auxiliary path contributes. The resulting operators flip the $0/2$ species at odd and even positions, respectively, in their ordered sequence, independently of the positions of the $1$'s. Since these transformations preserve the positions of the $0/2$ species, applying either transformation twice restores the original configuration. Hence,
\begin{align}
    &(\widetilde U^{(\nu)})^2 = \bm{1}, \nonumber \\
    &\widetilde U^{(\nu)}\big|_{W_{\rm emb}}
    = U^{(\nu)}, \qquad \nu=1,2.
\end{align}

The $XXC$ Hamiltonian preserves the order of the IS word and acts identically on the two species. Therefore, the transformations $\widetilde U^{(\nu)}$ commute with the unperturbed dynamics, 
\begin{align}
    [\widetilde U^{(\nu)},\,H_{XXC}] = 0.
\end{align}
In particular, $\widetilde U^{(\nu)}$ maps each IS sector $W_\lambda$ unitarily onto another IS sector $W_{f_\nu(\lambda)}$, where $f_\nu$ denotes the corresponding transformation of the IS word. The commutation relation implies
\begin{align}
H_{XXC}\big|_{W_{f_\nu(\lambda)}} \widetilde{U}^{(\nu)}\big|_{W_\lambda}
    &=
    \widetilde{U}^{(\nu)}\big|_{W_\lambda} H_{XXC}\big|_{W_\lambda}.
\end{align}
Thus, the restricted Hamiltonians on the two IS sectors are unitarily equivalent. The extension is not unique, since its action outside $W_{\rm emb}$ is not fixed by the restricted correspondence. 
Note that the existence of a full-space intertwiner for the unperturbed $XXC$ dynamics does not imply that conjugation maps arbitrary local perturbations to local operators. The local realization of the transformed embedded dynamics and leakage is instead established directly below.

We now give a local realization of the polarized-to-alternating correspondence for the embedded dynamics and leakage. Although the MPO constructed above intertwines the unperturbed $XXC$ dynamics, its conjugation does not generally preserve the locality of perturbations. Here we construct the corresponding local terms directly.

Let $P_w$ be the projector onto $w = {\rm span}\{|0\rangle,|2\rangle\}^{\otimes 2}$. The embedded density $h_w$ defined in \eqref{eq:decomp_h} obeys $h_w = P_w h_w P_w$. Under the staggered spin flip $U^{(\nu)}$ ($\nu = 1,2$) on $W_{\rm emb}$, the embedded density transforms by conjugation with a single spin flip on each bond. In particular,
\begin{align}
    &(\sigma^x\otimes\bm{1}) h_{XXZ}(\widetilde{a},\widetilde{b},\widetilde{c}) (\sigma^x\otimes\bm{1})
    = h_{XYZ}(\widetilde{b},\widetilde{a},0,\widetilde{c}), \nonumber \\
    &(\bm{1}\otimes\sigma^x) h_{XXZ}(\widetilde{a},\widetilde{b},\widetilde{c}) (\bm{1}\otimes\sigma^x)
    = h_{XYZ}(\widetilde{b},\widetilde{a},0,\widetilde{c}).
\end{align}
These identities establish the restricted unitary relation \eqref{eq:c_to_d} at the level of local densities. 

In fact, for the embedded perturbations considered here, the correspondence extends to the full Hilbert space. Since the perturbation acts only on adjacent $0/2$ pairs, which are also consecutive in the ordered IS word, the MPO $\widetilde{U}^{(\nu)}$ flips exactly one state on each such pair. 
Consequently,
\begin{align}
    \widetilde{U}^{(\nu)} H_{\rm pXXC}^{\rm pol} (\widetilde U^{(\nu)})^\dagger
    = H_{\rm pXXC}^{\rm alt},
\end{align}
with the parameter correspondence specified in \eqref{eq:c_to_d}. 

The leakage can also be transformed locally. For a leakage operator $h_{w \to w^{\bot}} = Q_w h_{w \to w^{\bot}} P_w$, with $Q_w = \bm{1} - P_w$, define
\begin{align}
    &(h_{w \to w^{\bot}}')_{j,j+1}
    = (h_{w \to w^{\bot}})_{j,j+1} (U_j^{(\nu)} \otimes U_{j+1}^{(\nu)})^\dagger, \nonumber \\
    &h'_{\rm mix} 
    = h_{w \to w^{\bot}}' + (h_{w \to w^{\bot}}')^\dagger,
\end{align}
where $U^{(\nu)}_j$ is the $j$th-site restriction of the staggered spin flip. 
Unlike the first reduction, the local realization of the second reduction does not rely on full-space unitary equivalence under the MPO transformation. For the antisymmetric leakage \eqref{eq:pol_leakage}, it reproduces staggered transformed local leakage operators \eqref{eq:neel_leakage} and the corresponding bondwise annihilation property.

The transformed embedded density is one-site uniform, whereas the transformed leakage alternates in sign between neighboring bonds. Both remain local, and the resulting Hamiltonian preserves the transformed tower.

%%%%%%%%%%%%%%%%%%%%%%%%%%%%%%%%%%%%%%%%%%%%%%%%%%%%%%%
%%%%%%%%%%%%%%%%%%%%%%%%%%%%%%%%%%%%%%%%%%%%%%%%%%%%%%%
\section{Local consistency of unitary realizations}
\label{app:local_realization}
%%%%%%%%%%%%%%%%%%%%%%%%%%%%%%%%%%%%%%%%%%%%%%%%%%%%%%%
%%%%%%%%%%%%%%%%%%%%%%%%%%%%%%%%%%%%%%%%%%%%%%%%%%%%%%%

Appendix~\ref{app:restricted_correspondence} gives an explicit local realization of the polarized-to-alternating correspondence for the embedded dynamics and leakage.
Here we consider periodic transformations within $W_{\rm emb}$ and examine when the transformed dynamics admits a one-site translationally invariant local realization with the same interaction range. We first consider transformations that flip exactly one site in each block of period $p$. Then multiple-site flips are discussed under an additional parity condition. 

We impose local consistency at the level of the Hamiltonian density, requiring the transformed local densities to be independent of their positions relative to the periodic pattern. This provides a sufficient condition for one-site translational invariance. More general realizations involving cancellations of telescoping terms are not considered here.

Let the perturbation restricted to $W_{\rm emb}$ be written as
\begin{align}
    H_W^{(r)}
    = \sum_{j=1}^{L-r+1} \left( h_W^{(r)} \right)_{j,j+1,\ldots,j+r-1}, \label{eq:local_HW}
\end{align}
where $h_W^{(r)}$ is supported on $r$ consecutive sites. Within $W_{\rm emb}$, we consider periodic block transformations generated by the local spin flip
\begin{align}
    U_j
    = \sigma_j^x, \label{eq:local_flip}
\end{align}
which exchanges the two effective spin states corresponding to $|0\rangle$ and $|2\rangle$. 

We first consider the staggered case corresponding to period $p=2$ for interaction range $r=2$. For even $L$, the two transformations are
\begin{align}
    U^{(1;2)} = \bigotimes_{j=1}^{L/2} U_{2j-1}, \qquad
    U^{(2;2)} = \bigotimes_{j=1}^{L/2} U_{2j}. \label{eq:period2_flip}
\end{align}
Every nearest-neighbor bond contains exactly one transformed site. Depending on the position of the bond relative to the staggered pattern, the spin flip acts either on the first or on the second site. Local consistency at the level of the Hamiltonian density then requires 
\begin{align}
    U_j ( h_W^{(2)} )_{j,j+1} U_j^\dagger
    = U_{j+1} ( h_W^{(2)} )_{j,j+1} U_{j+1}^\dagger
    \equiv h'^{(2)}_W. \label{eq:local_consistency_2}
\end{align}
Notice that local consistency itself does not require $h'^{(2)}_W = h_W^{(2)}$. Thus, a nontrivial transformation of the local dynamics can be compatible with one-site translational invariance.

Writing
\begin{align}
    h_W^{(2)}
    = \sum_{s,t,s',t'=\uparrow,\downarrow}
    a_{s,t}^{s',t'} |s,t\rangle\langle s',t'|, \label{eq:general_h2}
\end{align}
and denoting the flipped spin by $\bar{\uparrow} = \downarrow$ and $\bar{\downarrow} = \uparrow$, \eqref{eq:local_consistency_2} gives
\begin{align}
    a_{\bar{s},t}^{\bar{s}',t'}
    = a_{s,\bar{t}}^{s',\bar{t}'}. \label{eq:coeff_consistency_2}
\end{align}
The polarized-to-alternating transformation considered in Sec.~\ref{sec:pol_alt} satisfies this condition. It therefore provides a nontrivial nearest-neighbor realization within the one-site translationally invariant class.

We next consider a block period longer than the interaction range. The simplest case is $r = 2 < p$. In contrast to the staggered case with $p = 2$, translating a nearest-neighbor bond relative to the block pattern now produces both bonds containing a transformed site and bonds containing no transformed site. One-site translational invariance therefore requires the transformed local density to coincide with the original one. For the nearest-neighbor case,
\begin{align}
    U_\nu h_W^{(2)} U_\nu^\dagger
    = h_W^{(2)}, \qquad
    \nu = 1,2.
    \label{eq:r2_less_p}
\end{align}
Thus, when $r = 2 < p$, the obstruction is not to the unitary realization itself. Rather, local consistency requires invariance under each individual spin flip if the transformed dynamics is to remain within the same one-site translationally invariant nearest-neighbor Hamiltonian class. 

The same argument applies more generally when the interaction range is shorter than the block period, $r < p$. For a single flipped site per period and $r < p$, translating an $r$-site support produces at least one untransformed pattern and each of the $r$ possible single-site-flip patterns. Consequently,
\begin{align}
    U_\nu h_W^{(r)} U_\nu^\dagger
    = h_W^{(r)}, \qquad
    \nu = 1, \dots, r.
    \label{eq:r_less_p}
\end{align}
Hence, for a single flipped site per period with $r<p$, local consistency at the level of the local density forces invariance under each individual local spin flip. Note that this conclusion does not automatically extend to arbitrary multiple-site-flip patterns. 

This obstruction originates from the mismatch between the interaction range and the block period. When the interaction range is enlarged to match the block period, $r=p$, such that every translated $p$-site support contains exactly one transformed site, local consistency then requires
\begin{align}
    U_1 h_W^{(p)} U_1^\dagger
    = U_2 h_W^{(p)} U_2^\dagger
    = \cdots
    = U_p h_W^{(p)} U_p^\dagger
    \equiv h'^{(p)}_W, \label{eq:r_equal_p}
\end{align}
which does not require the invariance of the perturbation $h'^{(p)}_W = h_W^{(p)}$.

A simple nontrivial example for $p = r = 3$ is
\begin{align}
    h_W^{(3)}
    = a \sigma^z \otimes \sigma^z \otimes \sigma^z. \label{eq:p3_example}
\end{align}
Indeed, denoting by $U_\nu=\sigma_\nu^x$ the spin flip on the $\nu$th site of the three-site block, one finds
\begin{align}
    U_1 h_W^{(3)} U_1^\dagger
    = U_2 h_W^{(3)} U_2^\dagger
    = U_3 h_W^{(3)} U_3^\dagger
    = -h_W^{(3)}. \label{eq:p3_consistency}
\end{align}
Thus, a nontrivial transformed local density becomes possible again once the interaction range matches the block period.

This observation is also related to the period-three IS-preserving perturbations considered in \cite{bib:MT2024}. In the present notation, a period-three spin flip maps the fully polarized configuration to
$|\uparrow\uparrow\downarrow\uparrow\uparrow\downarrow\cdots \rangle$, corresponding through the natural embedding to the period-three IS $|002002\cdots \rangle$. The three-site perturbation constructed there preserves this period-three IS, although its generic form need not satisfy the stronger local consistency condition imposed here.

The local consistency condition \eqref{eq:r_equal_p} can also be expressed as
\begin{align}
    [ h_W^{(p)},\, U_\nu U_\mu ]
    = 0, \qquad
    \forall \nu,\mu. \label{eq:pair_flip_comm}
\end{align}
Using $U_\nu^2 = 1$ and the fact that transformations on different sites commute, it follows that for an arbitrary subset $\Lambda \subset \{ 1,\dots,p \}$,
\begin{align}
    U_\Lambda h_W^{(p)} U_\Lambda^\dagger
    =
    \begin{cases}
        h_W^{(p)}, & |\Lambda|\ {\rm even},\\
        h'^{(p)}_W, & |\Lambda|\ {\rm odd},
    \end{cases}
    \qquad
    U_\Lambda
    = \prod_{\nu \in \Lambda} U_\nu. \label{eq:parity_action}
\end{align}
Thus, when the single-flip consistency condition \eqref{eq:r_equal_p} holds, the transformation of the local density under multiple spin flips depends only on the parity of the number of flipped sites. If this parity is the same for all translated interaction supports, local consistency is satisfied. An even parity leaves the local density unchanged, whereas an odd parity allows a nontrivial transformation.

The results above show that local consistency depends on the relation between the transformation period $p$ and the interaction range $r$. For a single flipped site per period with $r<p$, local consistency requires invariance of the local Hamiltonian density under individual spin flips, whereas nontrivial transformations become possible already at $r=p$. The polarized-to-alternating construction provides an explicit example for $p=r=2$. These conditions are sufficient to ensure one-site translational invariance of the transformed Hamiltonian, without relying on cancellations of telescoping terms.

%%%%%%%%%%%%%%%%%%%%%%%%%%%%%%%%%%%%%%%%%%%%%%%%%%%%%%%
%%%%%%%%%%%%%%%%%%%%%%%%%%%%%%%%%%%%%%%%%%%%%%%%%%%%%%%
\section{Unitary realization of the spin-$1$ $XY$ scar model}
\label{app:XY}
%%%%%%%%%%%%%%%%%%%%%%%%%%%%%%%%%%%%%%%%%%%%%%%%%%%%%%%
%%%%%%%%%%%%%%%%%%%%%%%%%%%%%%%%%%%%%%%%%%%%%%%%%%%%%%%

In this appendix, we show explicitly how the spin-$1$ $XY$ scar model is obtained from the uniform $\mathfrak{su}(2)$ construction introduced in Sec.~\ref{sec:tower} through a site-dependent gauge transformation.

Recall that the scar tower of the spin-$1$ $XY$ model is generated by the staggered $\mathfrak{su}(2)$ generators~\cite{bib:SI2019,bib:OBCK2020,bib:M2026}
\begin{align}
    &T^{\pi;\pm}_{\rm tot}
    = V T^{\pm}_{\rm tot} V^{\dag}
    = \sum_{j=1}^L (-1)^j (S_j^{\pm})^2, \nonumber \\
    &T^{\pi;z}_{\rm tot}
    = V T^z_{\rm tot} V^{\dag}
    = \sum_{j=1}^L S_j^z
    = T^z_{\rm tot}, 
\end{align}
with the unitary gauge transformation given by $V = \otimes_{j=1}^L e^{i\frac{\pi j}{2}S_j^z}$. 

Since the fully polarized state $|\Uparrow\rangle$ is invariant under $V$ up to an overall phase, the uniform $\mathfrak{su}(2)$ tower introduced in \eqref{eq:ladder} is mapped to 
\begin{align}
    |\Psi_n^{{\rm pol};V}\rangle
    = V|\Psi_n^{\rm pol}\rangle 
    \propto (T^{\pi;-}_{\rm tot})^n|\Uparrow\rangle,
\end{align}
which spans the scar subspace $W^{(\pi)}_{\rm max}=V W_{\rm max}$ of the spin-$1$ $XY$ model. 
Moreover, the embedded spin-$1/2$ subspace is invariant under the transformation, $V W_{\rm emb} = W_{\rm emb}$. 

Accordingly, the mixing perturbation introduced in Sec.~\ref{sec:tower}, which breaks the invariance of $W_{\rm emb}$ while preserving the fully symmetric tower, is mapped to 
\begin{align}
    H_{\rm mix}(\pi)
    = V H_{\rm mix} V^\dagger
    = \sum_{j=1}^{L-1} (h_{\rm mix}(\pi))_{j,j+1},
\end{align}
where
\begin{align}
    h_{\rm mix}(\pi)
    = i(&\alpha |11 \rangle\langle 02|
    - \alpha^* |02 \rangle \langle 11| \nonumber \\
    + &\alpha |11 \rangle \langle 20|
    - \alpha^* |20 \rangle \langle 11|).
\end{align}
This coincides with the annihilation term of the spin-$1$ $XY$ scar model~\cite{bib:SI2019} up to the uniform transformation $G = \otimes_{j=1}^L {\rm diag}\,(1,1,i)$. 

By construction, $H_{\rm mix}(\pi)$ breaks the invariance of $W_{\rm emb}$ while annihilating the transformed scar subspace $W^{(\pi)}_{\rm max}$. Together with the Cartan term proportional to $T^{\pi;z}_{\rm tot}=T^z_{\rm tot}$, this gives the construction of the standard spin-$1$ $XY$ scar model~\cite{bib:SI2019}.

%%%%%%%%%%%%%%%%%%%%%%%%%%%%%%%%%%%%%%%%%%%%%%%%%%%%%%%
%%%%%%%%%%%%%%%%%%%%%%%%%%%%%%%%%%%%%%%%%%%%%%%%%%%%%%%
\section{Reduction hierarchy within the general $XXC$ family}
\label{app:general_xxc}
%%%%%%%%%%%%%%%%%%%%%%%%%%%%%%%%%%%%%%%%%%%%%%%%%%%%%%%
%%%%%%%%%%%%%%%%%%%%%%%%%%%%%%%%%%%%%%%%%%%%%%%%%%%%%%%

The spin-$1$ $XXC$ model studied in the main text is the simplest member of a broader family of $XXC$ models~\cite{bib:M1999}. In this appendix, we extend the two reduction steps to this general family and identify the structures that support the resulting hierarchy. 

Throughout the main text, we specialize to
\begin{align}
    A=\{0,2\}, \qquad
    B=\{1\},
\end{align}
which gives the spin-$1$ $XXC$ Hamiltonian considered in Sec.~\ref{sec:XXC}.
For an arbitrary partition $A\cup B = \{ 0,\dots,\ell \}$, the ordering of the local species inside each subset is conserved by the
general $XXC$ dynamics.
The direct generalization of the spin-$1$ embedded space is the all-$A$ subspace $W_{\rm emb} = {\rm span}\{ |s\rangle:s\in A \}^{\otimes L}$. It is a direct sum of frozen full-length IS sectors, not a subspace contained in each individual fragmented sector. The unperturbed general $XXC$ Hamiltonian \eqref{eq:gen_XXC} acts as a scalar on this space, 
\begin{align}
    H_{XXC}^{\rm gen}(a,b,c) \big|_{W_{\rm emb}} = a(L-1) \bm{1}, 
\end{align}
since every adjacent pair belongs to $A \times A$. Thus, as in the spin-$1$ example, local perturbations restricted to $W_{\rm emb}$ can couple different originally frozen IS words while preserving their collective span. 

Since $W_{\rm emb}$ is invariant under the unperturbed general $XXC$
Hamiltonian, arbitrary local dynamics may be introduced within
$W_{\rm emb}$ without destroying its invariance.
The resulting embedded dynamics need not be integrable. A representation-based further reduction requires an appropriate algebraic structure compatible with the chosen embedded dynamics; it is not implied by the existence of arbitrary local dynamics on $W_{\rm emb}$.

This construction realizes the first reduction step of the hierarchy,
namely the reduction from exponentially many fragmented invariant sectors
to one (or a few) exponentially large invariant subspaces.

The second reduction step requires a suitable representation structure in the dynamics introduced within $W_{\rm emb}$. A natural choice for the general $XXC$ family is to impose a global $\mathfrak{su}(|A|)$ symmetry on the embedded Hamiltonian, treating the local states belonging to $A$ as the fundamental representation. The embedded space then decomposes into invariant isotypic components under this symmetry. In particular, the fully symmetric representation occurs with multiplicity one and hence forms an invariant subspace.

The fully symmetric subspace is given by
\begin{align}
    W_{\rm sym}
    = {\rm Sym}^L(\mathbb{C}^{|A|})
    \subset W_{\rm emb},
\end{align}
whose dimension grows polynomially with $L$ for fixed $|A|$:
\begin{align}
    \dim W_{\rm sym}
    = \binom{L+|A|-1}{|A|-1}
    = O(L^{|A|-1}).
\end{align}
This provides a natural candidate for the final invariant subspace of the reduction hierarchy.

To preserve $W_{\rm sym}$ while breaking the invariance of $W_{\rm emb}$, we introduce a local coupling between $W_{\rm emb}$ and its complement that annihilates the fully symmetric states while acting nontrivially on other states. Let $\Pi_{j,j+1}$ denote the permutation operator exchanging the local $A$ degrees of freedom on neighboring sites. Since
\begin{align}
    (1 - \Pi_{j,j+1}) W_{\rm sym}
    = 0,
\end{align}
the antisymmetric component of the two-site $A$ subspace provides a natural source of such a coupling.

More explicitly, let us consider the mixing perturbation 
\begin{align}
    H_{\rm mix}
    &= \sum_{j=1}^{L-1}
    (h_{w \to w_{\bot}} + h^{\dag}_{w \to w_{\bot}})_{j,j+1}, 
\end{align}
consisting of the mixing term $(h_{w \to w_{\bot}})_{j,j+1}$ that maps the two-site $A$ subspace to its orthogonal complement, satisfying 
\begin{align}
    (h_{w \to w_{\bot}})_{j,j+1}
    &= (h_{w \to w_{\bot}})_{j,j+1}\, \frac{\bm{1} - \Pi_{j,j+1}}{2}.
\end{align}
This expression with the antisymmetric projector $(\bm{1} - \Pi_{j,j+1})/2$ guarantees that $(h_{w \to w_{\bot}})_{j,j+1}$ annihilates $W_{\rm sym}$. Since $(h_{w \to w_{\bot}})_{j,j+1}$ maps the two-site $A$ subspace to its orthogonal complement, its adjoint also annihilates $W_{\rm sym}$. Consequently, we have 
\begin{align}
    H_{\rm mix} W_{\rm sym}
    = 0,
\end{align}
whereas $H_{\rm mix}$ generally breaks the invariance of $W_{\rm emb}$.

For $|A|=2$ and $B=\{1\}$, the construction reduces to the antisymmetric leakage operator introduced in \eqref{eq:pol_leakage},
\begin{align}
    (h_{w \to w^{\bot}})_{j,j+1}
    &= \alpha \big( |11\rangle (\langle 02| - \langle 20|) \big)_{j,j+1}, 
\end{align}
which preserves the fully symmetric $\mathfrak{su}(2)$ tower while coupling other states in $W_{\rm emb}$ to its complement.

Since the fully symmetric representation occurs with multiplicity one, an embedded Hamiltonian with global $\mathfrak{su}(|A|)$ symmetry acts as a scalar on $W_{\rm sym}$, by Schur's lemma:
\begin{align}
    H_{\rm emb}\big|_{W_{\rm sym}}
    = E_{\rm sym}\bm{1}.
\end{align}
The resulting invariant subspace is therefore initially degenerate. A nontrivial energy structure can be introduced by adding a field proportional to the Cartan generators,
\begin{align}
    H_{\rm field}
    = \sum_{j=1}^L \sum_{a=1}^{|A|-1} h_a H_j^{(a)}.
\end{align}
Although this term generally breaks the global $\mathfrak{su}(|A|)$ symmetry, it preserves $W_{\rm sym}$. The Hamiltonian restricted to this subspace becomes
\begin{align}
    H\big|_{W_{\rm sym}}
    = \big(a(L-1) + E_{\rm sym}\big)\bm{1}
    + \sum_{a=1}^{|A|-1}  h_a H_{\rm tot}^{(a)},
\end{align}
where $H_{\rm tot}^{(a)} = \sum_j H_j^{(a)}$. The energy eigenstates within $W_{\rm sym}$ are therefore labeled by the weights of the fully symmetric representation, and generic choices of the coefficients $h_a$ lift their degeneracies.

This construction reproduces the representation-based scar mechanism discussed in \cite{bib:M2026}, now realized through successive reductions of the fragmented Hilbert space. The first reduction exploits the scalar action of the general $XXC$ Hamiltonian on $W_{\rm emb}$, allowing new embedded dynamics to be introduced. The second uses the representation structure of that dynamics and a compatible local coupling to preserve a polynomial-dimensional invariant subspace while breaking the invariance of $W_{\rm emb}$.

For the general $XXC$ family with $|A|\geq 2$ and $B \neq \varnothing$, a nonzero local coupling of this form can always be constructed by mapping a two-site antisymmetric state in $A \otimes A$ to a state outside $A \otimes A$. Thus, the fully symmetric $\mathfrak{su}(|A|)$ representation provides an explicit realization of the second reduction. Extensions to other representation structures or fragmentation mechanisms require a separate construction of compatible local couplings.

The general $XXC$ construction shows that the reduction hierarchy is not specific to the spin-$1$ realization. Its essential ingredients are the scalar action of the original Hamiltonian on an embedded space, the representation structure of the dynamics introduced within that space, and local couplings that preserve selected representation subspaces. 
The resulting invariant subspaces can further support representation-based scar spectra through the addition of Cartan generators~\cite{bib:OBCK2020,bib:M2026}.

The resulting hierarchy can be summarized as
\begin{align*}
    \underbrace{\text{$XXC$ fragmentation}}_{\substack{\text{exponentially many}\\ \text{IS sectors}}}
    \longrightarrow
    \underbrace{W_{\rm emb}}_{\substack{\text{exponential}\\ \text{dimension}}} 
    \longrightarrow
    \underbrace{W_{\rm sym}}_{\substack{\text{polynomial}\\ \text{dimension}}}.  
\end{align*}

\bibliography{references}

@article{bib:D1991,
  author   = {Deutsch, J. M.},
  title    = {Quantum statistical mechanics in a closed system},
  journal  = {Phys. Rev. A},
  volume   = {43},
  pages    = {2046--2049},
  year     = {1991},
  doi      = {10.1103/PhysRevA.43.2046}
}

@article{bib:S1994,
  author   = {Srednicki, M.},
  title    = {Chaos and quantum thermalization},
  journal  = {Phys. Rev. E},
  volume   = {50},
  pages    = {888--901},
  year     = {1994},
  doi      = {10.1103/PhysRevE.50.888}
}

@article{bib:S1999,
  author   = {Srednicki, M.},
  title    = {The approach to thermal equilibrium in quantized chaotic systems},
  journal  = {J. Phys. A: Math. Gen.},
  volume   = {32},
  pages    = {1163--1175},
  year     = {1999},
  doi      = {10.1088/0305-4470/32/7/007}
}

@article{bib:RDO2008,
  author   = {Rigol, M. and Dunjko, V. and Olshanii, M.},
  title    = {Thermalization and its mechanism for generic isolated quantum systems},
  journal  = {Nature},
  volume   = {452},
  pages    = {854--858},
  year     = {2008},
  doi      = {10.1038/nature06838}
}

@article{bib:DKPR2016,
  author   = {D'Alessio, L. and Kafri, Y. and Polkovnikov, A. and Rigol, M.},
  title    = {From quantum chaos and eigenstate thermalization to statistical mechanics and thermodynamics},
  journal  = {Adv. Phys.},
  volume   = {65},
  pages    = {239--362},
  year     = {2016},
  doi      = {10.1080/00018732.2016.1198134}
}

@incollection{bib:P2022,
  author   = {Papi{\'c}, Z.},
  title    = {Weak Ergodicity Breaking Through the Lens of Quantum Entanglement},
  booktitle= {Entanglement in Spin Chains: From Theory to Quantum Technology Applications},
  editor   = {Bayat, A. and Bose, S. and Johannesson, H.},
  publisher= {Springer International Publishing},
  address  = {Cham},
  pages    = {341--395},
  year     = {2022},
  doi      = {10.1007/978-3-031-03998-0_13}
}

@article{bib:MBR2022,
  author   = {Moudgalya, S. and Bernevig, B. A. and Regnault, N.},
  title    = {Quantum many-body scars and {Hilbert} space fragmentation: a review of exact results},
  journal  = {Rep. Prog. Phys.},
  volume   = {85},
  pages    = {086501},
  year     = {2022},
  doi      = {10.1088/1361-6633/ac73a0}
}

@article{bib:KHN2020,
  author   = {Khemani, V. and Hermele, M. and Nandkishore, R.},
  title    = {Localization from {Hilbert} space shattering: From theory to physical realizations},
  journal  = {Phys. Rev. B},
  volume   = {101},
  pages    = {174204},
  year     = {2020},
  doi      = {10.1103/PhysRevB.101.174204}
}

@article{bib:SRVKP2020,
  author   = {Sala, P. and Rakovszky, T. and Verresen, R. and Knap, M. and Pollmann, F.},
  title    = {Ergodicity Breaking Arising from {Hilbert} Space Fragmentation in Dipole-Conserving {Hamiltonians}},
  journal  = {Phys. Rev. X},
  volume   = {10},
  pages    = {011047},
  year     = {2020},
  doi      = {10.1103/PhysRevX.10.011047}
}

@article{bib:S1995,
  author  = {Sato, R.},
  title   = {{Bethe} ansatz calculations for quantum spin chains with partial integrability},
  journal = {J. Phys. Soc. Jpn.},
  volume  = {64},
  pages   = {2837--2847},
  year    = {1995},
  doi     = {10.1143/JPSJ.64.2837}
}

@article{bib:SA1996,
  author  = {Sato, R. and Akutsu, Y.},
  title   = {First-order phase transition in the {$S=1$} quantum spin chain with partial integrability},
  journal = {J. Phys. Soc. Jpn.},
  volume  = {65},
  pages   = {1885--1887},
  year    = {1996},
  doi     = {10.1143/JPSJ.65.1885}
}

@article{bib:M2024,
  author  = {Matsui, C.},
  title   = {Exactly solvable subspaces of nonintegrable spin chains with boundaries and quasiparticle interactions},
  journal = {Phys. Rev. B},
  volume  = {109},
  pages   = {104307},
  year    = {2024},
  doi     = {10.1103/PhysRevB.109.104307}
}

@article{bib:ADS2024,
  author  = {Aditya, S. and Dhar, D. and Sen, D.},
  title   = {Subspace-restricted thermalization in a correlated-hopping model with strong {Hilbert} space fragmentation characterized by irreducible strings},
  journal = {Phys. Rev. B},
  volume  = {110},
  pages   = {045418},
  year    = {2024},
  doi     = {10.1103/PhysRevB.110.045418}
}

@article{bib:MT2024,
  author   = {Matsui, C. and Tsuji, N.},
  title    = {Boundary dissipative spin chains with partial solvability inherited from system {Hamiltonians}},
  journal  = {Phys. Rev. B},
  volume   = {110},
  pages    = {224306},
  year     = {2024},
  doi      = {10.1103/PhysRevB.110.224306}
}

@article{bib:KMPP2025,
  author   = {Katsura, H. and Matsui, C. and Paletta, C. and Pozsgay, B.},
  title    = {Weak ergodicity breaking with isolated integrable sectors},
  journal  = {Phys. Rev. Research},
  volume   = {7},
  pages    = {023099},
  year     = {2025},
  doi      = {10.1103/PhysRevResearch.7.023099}
}

@article{bib:DB1993,
  author   = {Dhar, D. and Barma, M.},
  title    = {Conservation laws in stochastic deposition-evaporation models in one dimension},
  journal  = {Pramana},
  volume   = {41},
  pages    = {L193--L198},
  year     = {1993},
  doi      = {10.1007/BF02847591}
}

@article{bib:BD1994,
  author   = {Barma, M. and Dhar, D.},
  title    = {Slow Relaxation in a Model with Many Conservation Laws: Deposition and Evaporation of Trimers on a Line},
  journal  = {Phys. Rev. Lett.},
  volume   = {73},
  pages    = {2135--2138},
  year     = {1994},
  doi      = {10.1103/PhysRevLett.73.2135}
}

@article{bib:MBD1997,
  author   = {Menon, G. I. and Barma, M. and Dhar, D.},
  title    = {Conservation laws and integrability of a one-dimensional model of diffusing dimers},
  journal  = {J. Stat. Phys.},
  volume   = {86},
  pages    = {1237--1263},
  year     = {1997},
  doi      = {10.1007/BF02183622}
}

@article{bib:MRBR2018,
  author   = {Moudgalya, S. and Rachel, S. and Bernevig, B. A. and Regnault, N.},
  title    = {Exact excited states of nonintegrable models},
  journal  = {Phys. Rev. B},
  volume   = {98},
  pages    = {235155},
  year     = {2018},
  doi      = {10.1103/PhysRevB.98.235155}
}

@article{bib:MRB2018,
  author   = {Moudgalya, S. and Regnault, N. and Bernevig, B. A.},
  title    = {Entanglement of exact excited states of {Affleck}-{Kennedy}-{Lieb}-{Tasaki} models: Exact results, many-body scars, and violation of the strong eigenstate thermalization hypothesis},
  journal  = {Phys. Rev. B},
  volume   = {98},
  pages    = {235156},
  year     = {2018},
  doi      = {10.1103/PhysRevB.98.235156}
}

@article{bib:SI2019,
  author   = {Schecter, M. and Iadecola, T.},
  title    = {Weak Ergodicity Breaking and Quantum Many-Body Scars in Spin-1 {XY} Magnets},
  journal  = {Phys. Rev. Lett.},
  volume   = {123},
  pages    = {147201},
  year     = {2019},
  doi      = {10.1103/PhysRevLett.123.147201}
}

@article{bib:ISX2019,
  author   = {Iadecola, T. and Schecter, M. and Xu, S.},
  title    = {Quantum many-body scars from magnon condensation},
  journal  = {Phys. Rev. B},
  volume   = {100},
  pages    = {184312},
  year     = {2019},
  doi      = {10.1103/PhysRevB.100.184312}
}

@article{bib:OCMCCN2019,
  author   = {Ok, S. and Choo, K. and Mudry, C. and Castelnovo, C. and Chamon, C. and Neupert, T.},
  title    = {Topological many-body scar states in dimensions one, two, and three},
  journal  = {Phys. Rev. Research},
  volume   = {1},
  pages    = {033144},
  year     = {2019},
  doi      = {10.1103/PhysRevResearch.1.033144}
}

@article{bib:CIKM2023,
  author   = {Chandran, A. and Iadecola, T. and Khemani, V. and Moessner, R.},
  title    = {Quantum Many-Body Scars: A Quasiparticle Perspective},
  journal  = {Annu. Rev. Condens. Matter Phys.},
  volume   = {14},
  pages    = {443--469},
  year     = {2023},
  doi      = {10.1146/annurev-conmatphys-031620-101617}
}

@article{bib:TMASP2018,
  author   = {Turner, C. J. and Michailidis, A. A. and Abanin, D. A. and Serbyn, M. and Papi{\'c}, Z.},
  title    = {Weak ergodicity breaking from quantum many-body scars},
  journal  = {Nat. Phys.},
  volume   = {14},
  pages    = {745--749},
  year     = {2018},
  doi      = {10.1038/s41567-018-0137-5}
}

@article{bib:SAP2021,
  author   = {Serbyn, M. and Abanin, D. A. and Papi{\'c}, Z.},
  title    = {Quantum many-body scars and weak breaking of ergodicity},
  journal  = {Nat. Phys.},
  volume   = {17},
  pages    = {675--685},
  year     = {2021},
  doi      = {10.1038/s41567-021-01230-2}
}

@article{bib:MRB2020,
  author   = {Moudgalya, S. and Regnault, N. and Bernevig, B. A.},
  title    = {{$\eta$}-pairing in {Hubbard} models: From spectrum generating algebras to quantum many-body scars},
  journal  = {Phys. Rev. B},
  volume   = {102},
  pages    = {085140},
  year     = {2020},
  doi      = {10.1103/PhysRevB.102.085140}
}

@article{bib:LX2021,
  author   = {Langlett, C. M. and Xu, S.},
  title    = {{Hilbert} space fragmentation and exact scars of generalized {Fredkin} spin chains},
  journal  = {Phys. Rev. B},
  volume   = {103},
  pages    = {L220304},
  year     = {2021},
  doi      = {10.1103/PhysRevB.103.L220304}
}

@article{bib:YMP2026,
  author   = {Yang, F. and Magoni, M. and Pichler, H.},
  title    = {Constructing quantum many-body scars from weak {Hilbert} space fragmentation},
  journal  = {Phys. Rev. B},
  volume   = {113},
  pages    = {174307},
  year     = {2026},
  doi      = {10.1103/cpt2-kqdy}
}

@article{bib:OBCK2020,
  author   = {O'Dea, N. and Burnell, F. and Chandran, A. and Khemani, V.},
  title    = {From tunnels to towers: Quantum scars from {Lie} algebras and {$q$}-deformed {Lie} algebras},
  journal  = {Phys. Rev. Research},
  volume   = {2},
  pages    = {043305},
  year     = {2020},
  doi      = {10.1103/PhysRevResearch.2.043305}
}

@unpublished{bib:M2026,
  author        = {Matsui, C.},
  title         = {Scar subspaces stabilized by algebraic closure: Beyond equally-spaced spectra and exact solvability},
  year          = {2026},
  eprint        = {2604.11015},
  archivePrefix = {arXiv},
  primaryClass  = {cond-mat.stat-mech},
  doi           = {10.48550/arXiv.2604.11015}
}

@article{bib:M1998,
  author   = {Maassarani, Z.},
  title    = {The {XXC} models},
  journal  = {Phys. Lett. A},
  volume   = {244},
  pages    = {160--164},
  year     = {1998},
  doi      = {10.1016/S0375-9601(98)00322-3}
}

@article{bib:AM1998,
  author   = {Arnaudon, D. and Maassarani, Z.},
  title    = {Integrable open boundary conditions for {XXC} models},
  journal  = {J. High Energy Phys.},
  volume   = {1998},
  number   = {10},
  pages    = {024},
  year     = {1998},
  doi      = {10.1088/1126-6708/1998/10/024}
}

@article{bib:M1999,
  author   = {Maassarani, Z.},
  title    = {Multiplicity models},
  journal  = {Eur. Phys. J. B},
  volume   = {7},
  pages    = {627--633},
  year     = {1999},
  doi      = {10.1007/s100510050655}
}

@article{bib:S1988,
  author   = {Sklyanin, E. K.},
  title    = {Boundary conditions for integrable quantum systems},
  journal  = {J. Phys. A: Math. Gen.},
  volume   = {21},
  pages    = {2375},
  year     = {1988},
  doi      = {10.1088/0305-4470/21/10/015}
}

@article{bib:DG1994,
  author   = {de Vega, H. J. and Gonzalez-Ruiz, A.},
  title    = {Boundary {$K$}-matrices for the {XYZ}, {XXZ} and {XXX} spin chains},
  journal  = {J. Phys. A: Math. Gen.},
  volume   = {27},
  pages    = {6129},
  year     = {1994},
  doi      = {10.1088/0305-4470/27/18/021}
}

\end{document}